\documentclass{aastex631}
\usepackage{color,xcolor}
\usepackage{gensymb}
\usepackage{amsmath}
\usepackage{multirow}
\usepackage{amsfonts}

\providecommand{\Unit}[1]{\ensuremath{\mathrm{~#1}}} 
\providecommand{\pc}{\Unit{pc}}
\providecommand{\parallax}{\ensuremath{\varpi}}

\newcommand\gaia{\textit{Gaia~}}
\newcommand{\logg}{\ensuremath{\log g~}}
\newcommand{\ha}{\ensuremath{\rm H\alpha}}
\newcommand{\kmps}{\ensuremath{\mathrm{km\,s^{-1}}}}

\shorttitle{A long-period pre-ELM from LAMOST}
\shortauthors{Zhang et al.}

\graphicspath{{./}{figures/}}

\begin{document}

\title{A long-period pre-ELM system discovered from LAMOST medium-resolution 
survey}

\author[0000-0002-2419-6875]{Zhi-Xiang Zhang}
\affiliation{Department of Astronomy, Xiamen University, Xiamen, Fujian 361005, P.R. China}

\author[0000-0002-5630-7859]{Ling-Lin Zheng}
\affiliation{Department of Astronomy, Xiamen University, Xiamen, Fujian 361005, P.R. China}

\author[0000-0003-3137-1851]{Wei-Min Gu}
\affiliation{Department of Astronomy, Xiamen University, Xiamen, Fujian 361005, P.R. China}

\author[0000-0002-0771-2153]{Mouyuan Sun}
\affiliation{Department of Astronomy, Xiamen University, Xiamen, Fujian 361005, P.R. China}

\author[0000-0002-5839-6744]{Tuan Yi}
\affiliation{Department of Astronomy, Xiamen University, Xiamen, Fujian 361005, P.R. China}

\author[0000-0002-0349-7839]{Jian-Rong Shi}
\affiliation{Key Laboratory of Optical Astronomy, National Astronomical Observatories, Chinese Academy of Sciences, Beijing 100012, P.R. China}
\affiliation{School of Astronomy and Space Science, University of Chinese Academy of Sciences, Beijing 100049, P.R. China}

\author[0000-0003-3116-5038]{Song Wang}
\affiliation{Key Laboratory of Optical Astronomy, National Astronomical Observatories, Chinese Academy of Sciences, Beijing 100012, P.R. China}

\author{Zhong-Rui Bai}
\affiliation{Key Laboratory of Optical Astronomy, National Astronomical Observatories, Chinese Academy of Sciences, Beijing 100012, P.R. China}

\author[0000-0002-6617-5300]{Hao-Tong Zhang}
\affiliation{Key Laboratory of Optical Astronomy, National Astronomical Observatories, Chinese Academy of Sciences, Beijing 100012, P.R. China}

\author[0000-0003-1359-9908]{Wen-Yuan Cui}
\affiliation{Department of Physics, Hebei Normal University, Shijiazhuang 050024, P.R. China}

\author[0000-0003-4874-0369]{Junfeng Wang}
\affiliation{Department of Astronomy, Xiamen University, Xiamen, Fujian 361005, P.R. China}

\author[0000-0001-7349-4695]{Jianfeng Wu}
\affiliation{Department of Astronomy, Xiamen University, Xiamen, Fujian 361005, P.R. China}

\author[0000-0002-0584-8145]{Xiang-Dong Li}
\affiliation{Department of Astronomy, Nanjing University, Nanjing 210023, P.R. China}
\affiliation{Key laboratory of Modern Astronomy and Astrophysics (Nanjing University), Ministry of Education, Nanjing 210023, P.R. China}

\author[0000-0003-2506-6906]{Yong Shao}
\affiliation{Department of Astronomy, Nanjing University, Nanjing 210023, P.R. China}
\affiliation{Key laboratory of Modern Astronomy and Astrophysics (Nanjing University), Ministry of Education, Nanjing 210023, P.R. China}

\author[0000-0002-2310-0982]{Kai-Xing Lu}
\affiliation{Yunnan Observatories, Chinese Academy of Sciences, Kunming 650011, P.R. China}
\affiliation{Key Laboratory for the Structure and Evolution of Celestial Objects, Chinese Academy of Sciences, Kunming 650011, P.R. China}

\author[0000-0002-4740-3857]{Yu Bai}
\affiliation{Key Laboratory of Optical Astronomy, National Astronomical Observatories, Chinese Academy of Sciences, Beijing 100012, P.R. China}

\author[0000-0002-6647-3957]{Chunqian Li}
\affiliation{Key Laboratory of Optical Astronomy, National Astronomical Observatories, Chinese Academy of Sciences, Beijing 100012, P.R. China}
\affiliation{School of Astronomy and Space Science, University of Chinese Academy of Sciences, Beijing 100049, P.R. China}

\author[0000-0003-2896-7750]{Jin-Bo Fu}
\affiliation{Department of Astronomy, Xiamen University, Xiamen, Fujian 361005, P.R. China}

\author[0000-0002-2874-2706]{Jifeng Liu}
\affiliation{Key Laboratory of Optical Astronomy, National Astronomical Observatories, Chinese Academy of Sciences, Beijing 100012, P.R. China}
\affiliation{School of Astronomy and Space Science, University of Chinese Academy of Sciences, Beijing 100049, P.R. China}
\affiliation{WHU-NAOC Joint Center for Astronomy, Wuhan University, Wuhan 430072, P.R. China}

\correspondingauthor{Wei-Min Gu, Mouyuan Sun}

\email{guwm@xmu.edu.cn}\email{msun88@xmu.edu.cn}

\begin{abstract}

We present LAMOST~J041920.07+072545.4 (hereafter J0419), a close binary 
consisting of a bloated 
extremely low mass pre-white dwarf (pre-ELM WD)
and a compact object with an 
orbital period of 0.607189~days. The large-amplitude ellipsoidal 
variations and the evident Balmer and He~I emission lines suggest 
a filled Roche lobe and ongoing mass transfer.
No outburst events were detected in the 15 years of monitoring of 
J0419, indicating a very low mass transfer rate. The 
temperature of the pre-ELM, $T_\mathrm{eff} = 5793_{-133}^{+124}\,\rm K$,
is cooler than the known ELMs, but hotter than most CV donors.
Combining the mean density within the Roche lobe
and the radius constrained from our SED fitting, 
we obtain the mass of the pre-ELM, $M_1 = 0.176\pm 0.014\,M_\odot$. 
The joint fitting of light and radial velocity curves
yields an inclination angle of $i = 66.5_{-1.7}^{+1.4}$ degrees, 
corresponding to the compact object mass of $M_2 = 1.09\pm 0.05\,M_\odot$.
The very bloated pre-ELM has a smaller surface gravity 
($\log g = 3.9\pm 0.01$, $R_1 = 0.78 \pm 0.02\,R_\odot$) than the known 
ELMs or pre-ELMs. The temperature and the luminosity 
($L_\mathrm{bol} = 0.62_{-0.10}^{+0.11}\,L_\odot$) of J0419 are close 
to the main sequence, which makes the selection of such systems through 
the HR diagram inefficient. Based on the evolutionary model,  the 
relatively long period and small $\log g$ indicate that J0419 
could be close to the ``bifurcation period'' in the orbit evolution, 
which makes J0419 to be a unique source to connect ELM/pre-ELM WD systems, 
wide binaries and cataclysmic variables. 

\end{abstract}


\section{Introduction} \label{sec:intro}

Extremely low mass white dwarfs (ELM WDs) are helium-core WDs with masses 
below $0.3\,M_\odot$ \citep{Li2019}, which are different from 
most WDs that have C/O cores with mass around $0.6\,M_\odot$ 
\citep{Kepler2015}. ELM WDs are thought to be born in interactive binaries 
and have lost most of their mass to the companions through 
either the stable Roche lobe overflow or the common-envelope 
evolution (CE), as the formation time of a WD with mass less than 
0.3\,$M_\odot$ produced from a single star exceeds the Hubble timescale.
\citet{chen2017,sun2018,Li2019} theoretical studied the formation of ELMs 
and showed that the progenitors of ELMs fill the Roche lobes 
at the end of the main sequence (MS) or near the base of the 
red giant branch (RGB). When the mass transfer ceases, 
possibly due to the stop of the magnetic braking driven orbital 
contraction \citep[see][]{sun2018,Li2019}, a pre-ELM with a helium 
core and a hydrogen envelope is formed. 
In this paper, we use pre-ELMs to refer to all progenitors of ELMs.
After the detachment of the binary, 
the envelope will continue burning to keep a nearly constant luminosity until
the burnable hydrogen is exhausted, and the radius of the envelope 
will gradually shrink. The hydrogen exhausted pre-ELM will enter the WD 
cooling track. 

The research of ELMs/pre-ELMs has gradually become active in recent years.
Many pre-ELMs or ELMs have pulsations 
\citep{Maxted2011,Maxted2013,Maxted2014a,Gianninas2016,zhang2017}, which 
provide unprecedented opportunities to explore their interiors.
The high accretion rate in the early stages of the ELM formation may 
contribute enough mass to the C/O WD companion in the binary
and makes it a progenitor of a Type Ia supernova \citep{Han2004}. 
The compact ELMs, such as J0651+2844, that has a period of 765\,s 
\citep{brown2011,Amaro2012}, could be resolved by future space-based 
gravitational-wave detectors \citep{Amaro2012,luo2016b}. 

More than 100 ELM WDs and their progenitors have been reported by several 
surveys, e.g., the Kepler project 
\citep{van2010,Carter2011,Breton2012,Rappaport2015}, 
the WASP project \citep{Maxted2011,Maxted2013,Maxted2014a,Maxted2014b}, the 
ELM survey 
\citep{brown2010,brown2012,brown2013,brown2016,brown2020,Kilic2011,Kilic2012,Kilic2015,Gianninas2014,Gianninas2015}, 
and the ELM survey South \citep{Kosakowski2020}. Most of the objects 
reported by previous works are ELMs in double degenerates 
(DD) binaries, and their companions are WDs \citep{Li2019} or neutron 
stars \citep{Istrate2014b,Istrate2014}. Some works reported the pre-ELMs 
in EL CVn-type binaries, which are post-mass transfer 
eclipsing binaries that are composed of an A/F-type main sequence star and 
a progenitor of ELM in the shrinking stage 
\citep{Maxted2011,Maxted2014a,wangkun2018,wangkun2020}. All of these 
sources have ended their mass interactions.

The pre-ELMs in mass transfer or terminated mass transfer recently have 
temperatures similar to the main sequence A and F stars and can not be 
selected by their colors. By inspecting light curves with large amplitude
ellipsoidal variability and luminosities below the main sequence, 
\citet{elbadry2021b,Badry2021} reported a sample of pre-ELMs in DDs with 
periods less than 6 hours. They name these sources proto-ELMs. 
The objects of \citet{elbadry2021b,Badry2021} have lower temperatures than 
ELMs and the pre-ELMs in EL CVns. Moreover, their objects with 
temperatures lower than 6500\,K have emission lines, and the 
rest objects with higher temperatures do not have, indicating that 
the sample of \citet{elbadry2021b,Badry2021} are in the transition from 
mass transfer to detached.

The pre-ELM WDs with stable mass transfer behave as cataclysmic variables 
(CVs). Unlike normal CVs, pre-ELMs are evolved stars with Helium-cores. 
They have much lower mass transfer rates than normal CVs, and do not 
show typical CV characteristics in the light curves, such as random 
variation in a short timescale, outburst events.
Normal CVs generally have small-mass donors in the main sequence 
whose orbital periods are several hours \citep{Knigge2006,Knigge2011}.
The stellar parameters, such as mass, radius, spectral type, and luminosity, 
are closely related to the orbital period, which is called ``donor sequence''
\citep{Patterson1984,Beuermann1998,Smith1998,Knigge2006,Knigge2011}.  
With bloated radius and helium cores, the evolved donors significantly 
deviate from the donor sequence. They have smaller mass and 
higher temperatures compared with the donor in normal CVs 
\citep{Podsiadlowski2003,van2005,Kalomeni2016}.

The evolution trace of the ELMs mainly depends upon the initial 
period and initial mass \citep{Li2019}. The donors with longer 
initial periods will be more evolved before the mass transfer, 
resulting in pre-ELMs with more bloated radii and longer periods. 
These long-period pre-ELMs are close to the main sequence and, 
therefore, cannot be selected using the HR diagram. Meanwhile, 
some long-period pre-ELMs may have periods  
close to the bifurcation period. Theoretically speaking,
for the systems with orbital periods longer than the 
bifurcation period (16--22 h), 
the donors ascend the giant branch as the mass transfer begins, and 
the systems evolve toward long orbital periods with mass loss 
\citep{Podsiadlowski2003}. For the 
systems whose period is shorter than the bifurcation period, 
the orbits of these targets are contracting rather than 
expanding because of 
magnetic braking. The pre-ELMs with periods close to the bifurcation period 
are special cases between these two situations and vital for our 
understanding of the evolution of ELM systems.

In this work, we report the discovery of a pre-ELM with a period of 14.6 
hours, which is much longer than typical pre-ELMs.
The orbital period of this source is about three times that of the sample in 
\citet{elbadry2021b}, so the surface gravity is less than that of all ELMs 
or pre-ELMs we have known. Because of the larger radius and higher
luminosity, this object almost falls on the main sequence, making it 
inefficient to select this type of object using the HR diagram. Thanks to the 
time-domain spectroscopic (e.g., the Large Sky Area Multi-Object Fiber 
Spectroscopic Telescope; LAMOST; see \citealt{cui2012,zhao2012}) and 
photometric surveys, we are able to select such a particular pre-ELM.

The paper is organized as follows. In Section \ref{sec:data}, we describe 
the data, which include the spectroscopic data from several telescopes or 
instruments, and the photometric data from publicly available photometric 
surveys. In Section \ref{sec:data_analysis}, we present the process 
of data measurement and analysis, including determination of orbital 
period, radial-velocity (RV) measurements, SED fitting, and spectral matching. Discussion 
and summary are made in Sections 4 and 5.

\section{Data}
\label{sec:data}

J0419 (R.A. = $04^h19^m20^s.07$, Decl. = $07\degree25'45''.4$, J2000) is 
selected from the LAMOST medium-resolution surveys 
\citep[MRS;][]{liuchao2020} 
and has a stellar type of G8 and a magnitude of 14.70~mag in 
the \gaia $G$-band. The RV measurements of 
LAMOST DR8 MRS show that this source has an RV variation of about 212\,\kmps\ by 
six exposures on Nov 8, 2019. Since the absorption lines of the LAMOST spectra 
are single-lined, we speculate J0419 is a binary composed of a visible 
star and a compact object. We applied for additional 
spectroscopic observations to constrain the RV amplitude of J0419 by using 
the 2.16-meter telescope in Xinglong and the Lijiang 2.4-meter 
telescope. We also requested several LAMOST follow-up observations 
on this source. The observation information is summarized in Table 
\ref{tab:spec_stat}. In addition to spectroscopic data, we collected  
photometric data from several publicly available sky surveys, 
which include the Transiting Exoplanet Survey Satellite 
\citep[TESS;][]{Ricker2015}, the Catalina Real-time Transient Survey 
\citep[CRTS;][]{Drake2009,Drake2014}, the All-Sky Automated Survey for 
Supernovae \citep[ASAS-SN][]{Shappee2014,Kochanek2017} and the Zwicky 
Transient Facility \citep[ZTF;][]{Masci2019}. The data are described 
below.

\begin{deluxetable*}{llccccrrrr}
\tablenum{1}
\tablecaption{Statistics of the observed spectra of J0419}
\label{tab:spec_stat}
\tablewidth{0pt}
\tablehead{
\colhead{Num} &
\colhead{Telescope} & \colhead{HMJD} & \colhead{Obs. Date} & \colhead{Exp. time (s)} & \colhead{Phase} & \colhead{SNR} & \colhead{Resolution} & \colhead{RV (\kmps)} & \colhead{$\rm EW_{H\alpha}$ (\AA)}
}
\startdata
1  & LAMOST MRS    & 58795.69 & 2019-11-08 16:36:34 & 1200 & 0.93 & 10.1  & 7500 & $166.3_{-5.8}^{+5.8}$    &  $4.03\pm 0.17$ \\
2  & LAMOST MRS    & 58795.71 & 2019-11-08 16:59:34 & 1200 & 0.96 & 9.2   & 7500 & $129.7_{-4.8}^{+6.8}$    &  $3.41\pm 0.18$ \\
3  & LAMOST MRS    & 58795.72 & 2019-11-08 17:22:34 & 1200 & 0.99 & 8.4   & 7500 & $97.2_{-7.5}^{+5.8}$     &  $4.12\pm 0.20$ \\
4  & LAMOST MRS    & 58795.76 & 2019-11-08 18:12:34 & 1200 & 0.04 & 10.7  & 7500 & $17.5_{-5.8}^{+5.5}$     &  $4.36\pm 0.15$ \\
5  & LAMOST MRS    & 58795.78 & 2019-11-08 18:36:34 & 1200 & 0.07 & 10.1  & 7500 & $-13.3_{-4.0}^{+4.0}$    &  $4.22\pm 0.17$ \\
6  & LAMOST MRS    & 58795.79 & 2019-11-08 18:59:34 & 1200 & 0.10 & 10.3  & 7500 & $-45.8_{-4.8}^{+5.5}$    &  $3.94\pm 0.17$ \\
7  & LAMOST LRS    & 58837.61 & 2019-12-20 14:42:16 & 600  & 0.97 & 20.6  & 1800 & $104.5_{-7.0}^{+6.0}$    & $10.10\pm 0.17$ \\
8  & LAMOST LRS    & 58837.62 & 2019-12-20 14:56:16 & 600  & 0.99 & 20.2  & 1800 & $90.5_{-8.0}^{+7.0}$     &  $9.05\pm 0.17$ \\
9  & LAMOST LRS    & 58837.63 & 2019-12-20 15:09:16 & 600  & 0.00 & 21.5  & 1800 & $65.8_{-7.0}^{+7.0}$     &  $8.43\pm 0.16$ \\
10 & LAMOST LRS    & 58837.65 & 2019-12-20 15:31:16 & 600  & 0.03 & 23.5  & 1800 & $30.0_{-6.0}^{+6.0}$     &  $9.59\pm 0.16$ \\
11 & LAMOST LRS    & 58837.66 & 2019-12-20 15:45:16 & 600  & 0.05 & 25.2  & 1800 & $9.3_{-6.0}^{+6.0}$      &  $9.14\pm 0.15$ \\
12 & LAMOST LRS    & 58837.67 & 2019-12-20 15:58:16 & 600  & 0.06 & 26.3  & 1800 & $-12.8_{-6.8}^{+6.0}$    &  $9.73\pm 0.14$ \\
13 & 2.16-meter    & 59140.74 & 2020-10-18 17:39:32 & 1800 & 0.20 & 119.0 & 300  & -                        & $15.69\pm 0.12$ \\
14 & 2.16-meter    & 59140.76 & 2020-10-18 18:09:37 & 1800 & 0.23 & 107.7 & 300  & -                        & $15.64\pm 0.11$ \\
15 & 2.16-meter    & 59140.79 & 2020-10-18 18:56:15 & 1800 & 0.28 & 107.2 & 300  & -                        & $16.03\pm 0.11$ \\
16 & 2.16-meter    & 59140.81 & 2020-10-18 19:26:20 & 1800 & 0.32 & 102.4 & 300  & -                        & $16.65\pm 0.11$ \\
17 & 2.16-meter    & 59140.83 & 2020-10-18 19:57:45 & 1200 & 0.35 & 80.3  & 300  &  -                       & $16.22\pm 0.13$ \\
18 & 2.16-meter    & 59140.85 & 2020-10-18 20:17:50 & 1200 & 0.38 & 73.4  & 300  & -                        & $16.45\pm 0.14$ \\
19 & 2.16-meter    & 59140.86 & 2020-10-18 20:37:55 & 1200 & 0.40 & 70.9  & 300  & -                        & $17.03\pm 0.14$ \\
20 & 2.16-meter    & 59141.75 & 2020-10-19 17:56:03 & 1200 & 0.86 & 68.7  & 300  & -                        & $14.09\pm 0.17$ \\
21 & 2.16-meter    & 59141.76 & 2020-10-19 18:16:08 & 1200 & 0.88 & 67.7  & 300  & -                        & $13.36\pm 0.18$ \\
22 & 2.16-meter    & 59141.78 & 2020-10-19 18:36:13 & 1200 & 0.91 & 64.9  & 300  & -                        & $13.05\pm 0.18$ \\
23 & 2.16-meter    & 59141.80 & 2020-10-19 19:09:25 & 1200 & 0.95 & 64.1  & 300  & -                        & $12.07\pm 0.18$ \\
24 & 2.16-meter    & 59141.81 & 2020-10-19 19:29:30 & 1200 & 0.97 & 65.3  & 300  & -                        & $12.39\pm 0.17$ \\
25 & 2.16-meter    & 59141.83 & 2020-10-19 19:49:35 & 1200 & 0.99 & 63.3  & 300  & -                        & $14.59\pm 0.17$ \\
26 & 2.16-meter    & 59141.84 & 2020-10-19 20:10:01 & 1200 & 0.02 & 61.1  & 300  & -                        & $14.35\pm 0.19$ \\
27 & 2.16-meter    & 59141.85 & 2020-10-19 20:30:06 & 1200 & 0.04 & 61.2  & 300  & -                        & $15.92\pm 0.18$ \\
28 & 2.16-meter    & 59141.87 & 2020-10-19 20:50:11 & 1200 & 0.06 & 57.7  & 300  & -                        & $14.97\pm 0.17$ \\
29 & 2.16-meter    & 59192.62 & 2020-12-09 14:46:53 & 1800 & 0.64 & 47.4  & 620  & $232.5_{-24.0}^{+12.8}$  & $8.27\pm 0.20$  \\
30 & 2.16-meter    & 59192.64 & 2020-12-09 15:16:59 & 1800 & 0.67 & 48.9  & 620  & $284.2_{-24.0}^{+13.0}$  & $9.31\pm 0.21$  \\
31 & 2.16-meter    & 59192.66 & 2020-12-09 15:57:02 & 1800 & 0.72 & 56.1  & 620  & $318.7_{-19.8}^{+5.9}$   & $9.63\pm 0.17$  \\
32 & 2.16-meter    & 59192.69 & 2020-12-09 16:27:08 & 1800 & 0.75 & 56.3  & 620  & $303.0_{-21.2}^{+6.5}$   & $9.85\pm 0.17$  \\
33 & 2.16-meter    & 59192.71 & 2020-12-09 17:06:44 & 1800 & 0.80 & 55.7  & 620  & $278.8_{-21.9}^{+6.1}$   & $8.57\pm 0.17$  \\
34 & 2.16-meter    & 59192.74 & 2020-12-09 17:40:42 & 1800 & 0.84 & 53.2  & 620  & $265.1_{-22.3}^{+6.4}$   & $7.94\pm 0.19$  \\
35 & LAMOST MRS    & 59213.57 & 2020-12-30 13:37:59 & 1200 & 0.15 & 3.4   & 7500 & $-74.7_{-23.2}^{+21.2}$  & $4.83\pm 0.76$  \\
36 & LAMOST MRS    & 59213.58 & 2020-12-30 14:01:23 & 1200 & 0.17 & 3.5   & 7500 & $-66.6_{-21.2}^{+21.2}$  & $8.03\pm 1.10$  \\
37 & LAMOST MRS    & 59213.60 & 2020-12-30 14:24:45 & 1200 & 0.20 & 3.6   & 7500 & $-111.0_{-14.2}^{+14.1}$ & $11.95\pm 0.92$ \\
38 & LAMOST MRS    & 59213.62 & 2020-12-30 14:48:09 & 1200 & 0.23 & 3.7   & 7500 & $-118.1_{-18.1}^{+13.2}$ & $12.73\pm 0.75$ \\
39 & LAMOST MRS    & 59213.64 & 2020-12-30 15:18:54 & 1200 & 0.26 & 3.2   & 7500 & $-108.9_{-20.3}^{+27.2}$ & $13.20\pm 0.88$ \\
40 & LAMOST MRS    & 59213.65 & 2020-12-30 15:42:17 & 1200 & 0.29 & 3.1   & 7500 & $-112.0_{-67.6}^{+29.3}$ &          -      \\
41 & LAMOST MRS    & 59240.47 & 2021-01-26 11:13:44 & 1200 & 0.45 & 2.0   & 7500 & $7.0_{-18.3}^{+21.8}$    & $-0.19\pm 0.96$ \\
42 & LAMOST MRS    & 59240.48 & 2021-01-26 11:37:44 & 1200 & 0.48 & 2.6   & 7500 & $27.0_{-20.0}^{+24.8}$   &  $1.40\pm 0.83$ \\
43 & LAMOST MRS    & 59240.50 & 2021-01-26 12:00:44 & 1200 & 0.50 & 2.6   & 7500 & $98.7_{-56.1}^{+23.5}$   & $-0.22\pm 0.83$ \\
44 & LAMOST MRS    & 59242.47 & 2021-01-28 11:11:18 & 1200 & 0.74 & 5.8   & 7500 & $307.5_{-9.1}^{+10.0}$   &  $9.03\pm 0.60$ \\
45 & LAMOST MRS    & 59242.48 & 2021-01-28 11:34:40 & 1200 & 0.77 & 4.8   & 7500 & $305.4_{-24.1}^{+23.2}$  &  $8.13\pm 0.59$ \\
46 & LAMOST MRS    & 59242.50 & 2021-01-28 11:58:02 & 1200 & 0.79 & 5.0   & 7500 & $301.4_{-16.1}^{+13.1}$  &  $5.88\pm 0.58$ \\
47 & Lijiang 2.4-meter & 59248.55& 2021-02-03 13:09:16& 1801& 0.76& 40.5  & 850  & $299.7_{-12.0}^{+13.0}$  &  $2.67\pm 0.18$ \\
\enddata
\tablecomments{
The HMJD is the mid-exposure time.
The heliocentric corrections have been applied to the RVs.
We did not measure the RVs of the spectra observed by using the G4 
grism due to their low resolution. The spectrum of line 40 has no red arm 
data and therefore no information about the \ha\ emission line.
}
\end{deluxetable*}

\subsection{Spectroscopic Data}

\subsubsection{LAMOST spectra}

LAMOST is a uniquely designed 4-meter reflecting Schmidt  telescope that 
enables it to observe 4000 spectra simultaneously in a field of view of 
$5^{\circ}$ \citep{cui2012,zhao2012}. The wavelength coverage of LAMOST 
low-resolution ($R \sim 1800$) spectra ranges from 3690\,\AA\ to 9100\,\AA\ 
\citep{luo2016}. The LAMOST medium-resolution ($R \sim 7500$; see 
\citealt{liuchao2020}) spectra have two 
arms, of which the blue arm covers the wavelength range of  4950\,\AA\ to 
5350\,\AA, and the red arm covers a wavelength range of 6300\,\AA\ to 
6800\,\AA\ \citep{zong2018}. For both low- and medium-resolution spectra, 
the LAMOST's observation strategy is to perform 2--4 consecutive short 
exposures for 10--20 minutes each (see Table \ref{tab:spec_stat}). In the 
study of close binaries with a period of less than one day, the RV changes 
significantly between two single LAMOST exposures. Hence, the RVs of 
the short exposure LAMOST spectra are crucial in our study \citep{Mu2022}.

LAMOST MRS conducted the first observation of J0419 on Nov 8, 2019, with six 
consecutive exposures. The RVs (see Section \ref{sec:rv}) span from 
166\,\kmps\ to -46\,\kmps\ in 2.4 hours. LAMOST LRS made another six 
consecutive exposures on Dec 20, 2019, each with an exposure time of 
600\,s, and the resulting RVs are from 105\,\kmps\ to -13\,\kmps. 
On 2020-12-30, 2021-01-26, and 2021-01-28, LAMOST MRS performed 
follow-up observations on J0419 and obtained a total of 12 single exposure 
spectra. Due to the bright moon nights, the LAMOST follow-up spectra 
have very low SNRs. Nevertheless, we still use them to measure the 
corresponding RVs. The information of LAMOST spectroscopic data are 
summarized in Table \ref{tab:spec_stat}.

\begin{figure*}
    \centering
    \includegraphics[width=0.8\textwidth]{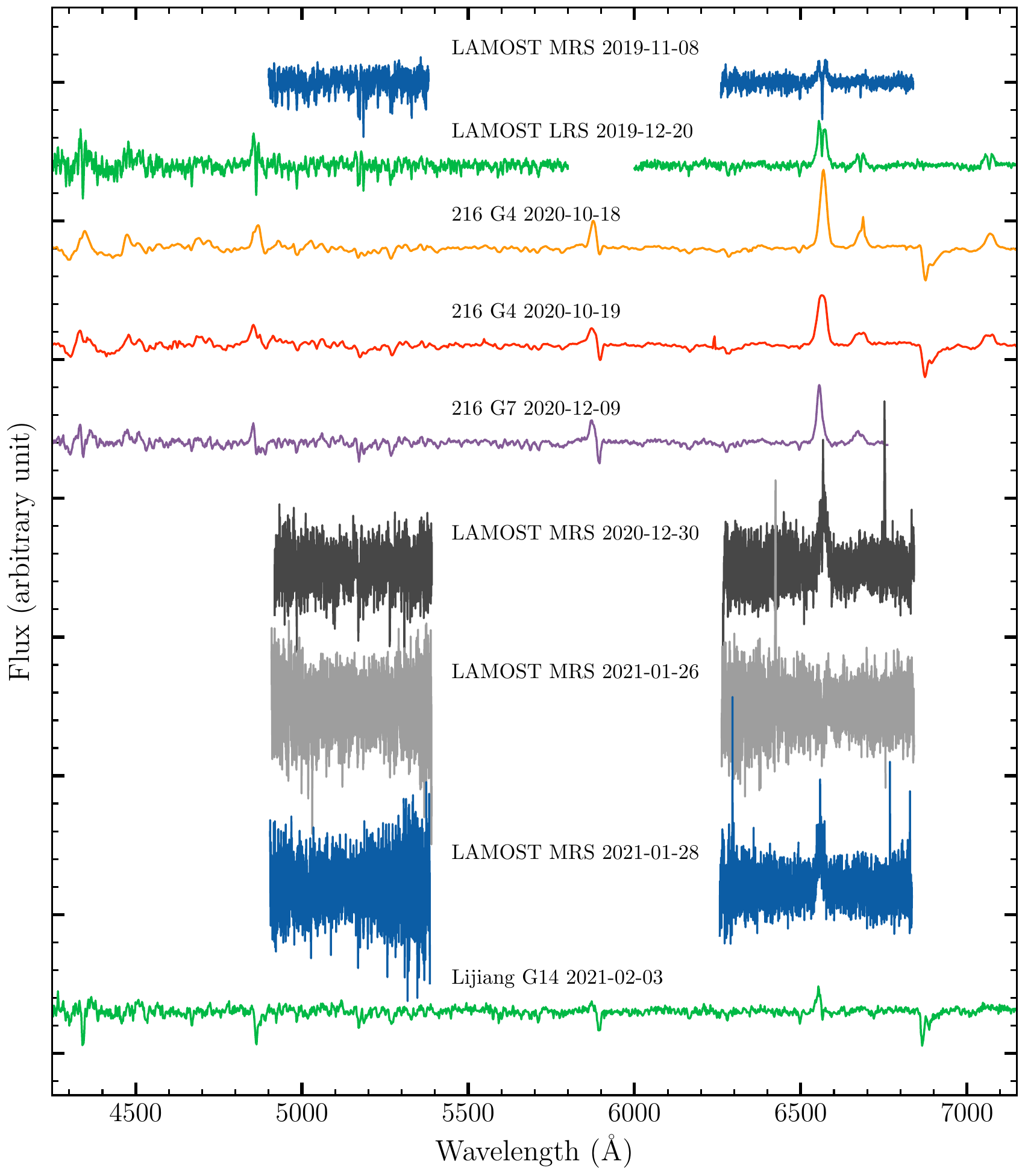}
    \caption{Normalized average spectra of J0419. Each average spectrum is 
    generated by combining the spectra observed on the same night. 
    Prior to 
    the combination, the wavelength has been shifted to rest frame. The 
    observation information is marked next to each average spectrum.}
    \label{fig:spec_all}
\end{figure*}

We combine the LAMOST spectra observed on the same night after correcting 
the wavelength of each spectrum to rest frame and plot them in Figure 
\ref{fig:spec_all}. The spectra show evident emission lines of 
Balmer and $\mathrm{He~I}$ with significant double peak characteristics 
in most of LAMOST observations, suggesting that the emission lines 
are not produced by the visible star. We discuss the 
emission lines in Section \ref{sec:dis_emission}.

\subsubsection{The 2.16-meter telescope spectra}

We applied for two spectral observations of J0419 using the 2.16-meter 
telescope \citep{Fan2016} at the Xinglong Observatory\footnote{\url{http://www.xinglong-naoc.org/}}.
The first observations were performed from 2020-10-18 to 2020-10-19, with 
16 exposures.
We chose the G4 grism and 1.8 slit combination, which yielded an 
instrumental broadening (FWHM) of 18.4\,\AA\ measured from skylines. The 
resolution is too low to measure the RVs. These spectra show strong Balmer 
and He~I emission lines (see Figure \ref{fig:spec_all}). 

The second observations were performed on Dec 9, 2020, and the grism was 
adjusted to G7 to improve the spectral resolution, which yielded an FWHM of 
9.0\,\AA\ measured from skylines. 
The spectra also show evident emission lines, albeit the 
equivalent widths (EW) of the emission lines are less than the first 
observations.

We used \texttt{IRAF v2.16} to reduce the spectra with standard process. 
The heliocentric correction was made using the \texttt{helcorr} function 
in python package \texttt{PyAstronomy}.

\subsubsection{The Lijiang 2.4-meter telescope spectra}
On February 3, 2021, we used the Yunnan Faint Object Spectrograph and Camera 
(YFOSC), which is mounted on the Lijiang 2.4-meter 
telescope\footnote{\url{http://gmg.org.cn/v2/}} at the Yunnan 
Observatories of the Chinese Academy of Sciences, to observe J0419. YFOSC is 
a multifunctional instrument both for photometry and spectroscopy that has a 
$2k \times 4k$ back-illuminated CCD detector. More information about YFOSC 
can be found in \citet{lu2019}. A grism G14 and a 1\arcsec.0 slit are used, 
resulting in wavelength coverage of 3800\,\AA\ -- 7200\,\AA\ with a 
spectral 
resolution of 6.5\,\AA\ measured from skylines. The Lijiang spectrum shows 
weak Balmer emission lines, and most of the He~I emissions cannot even be 
seen in the spectrum. The data reduction process of the Lijiang data is 
similar to that of the Xinglong 2.16-meter spectra.

\subsection{Photometric data}

\begin{figure*}
    \centering
    \includegraphics[width=\textwidth]{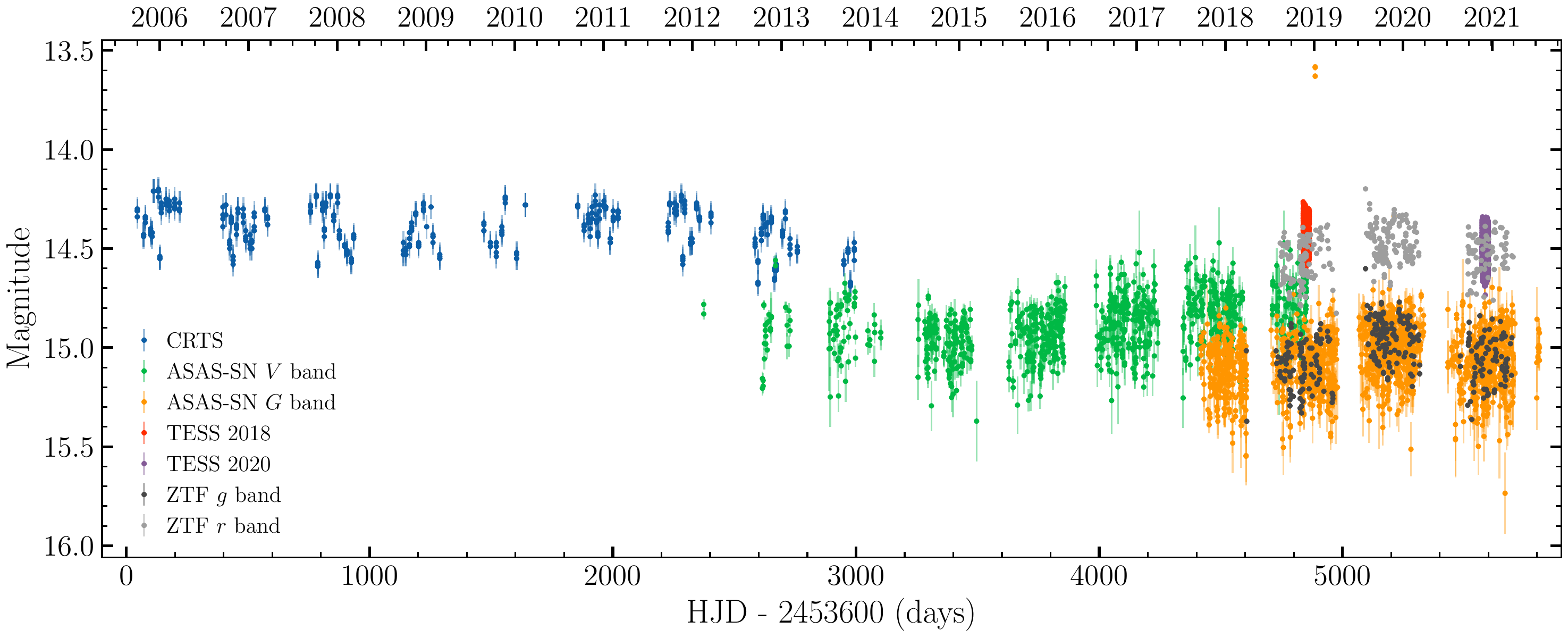}
    \caption{The light curves of J0419. Colors represent different surveys.
    No outburst events were captured 
    on the light curves.}
    \label{fig:lc_all}
\end{figure*}

We collect the light curves of J0419 from several publicly available 
photometric surveys. The light curves are used to determine the orbital 
period and analysis the variability (Figure \ref{fig:lc_all}).
We introduce the photometric data below.

\subsubsection{TESS}

\begin{figure}[htbp]
    \centering
    \includegraphics[width=0.47\textwidth]{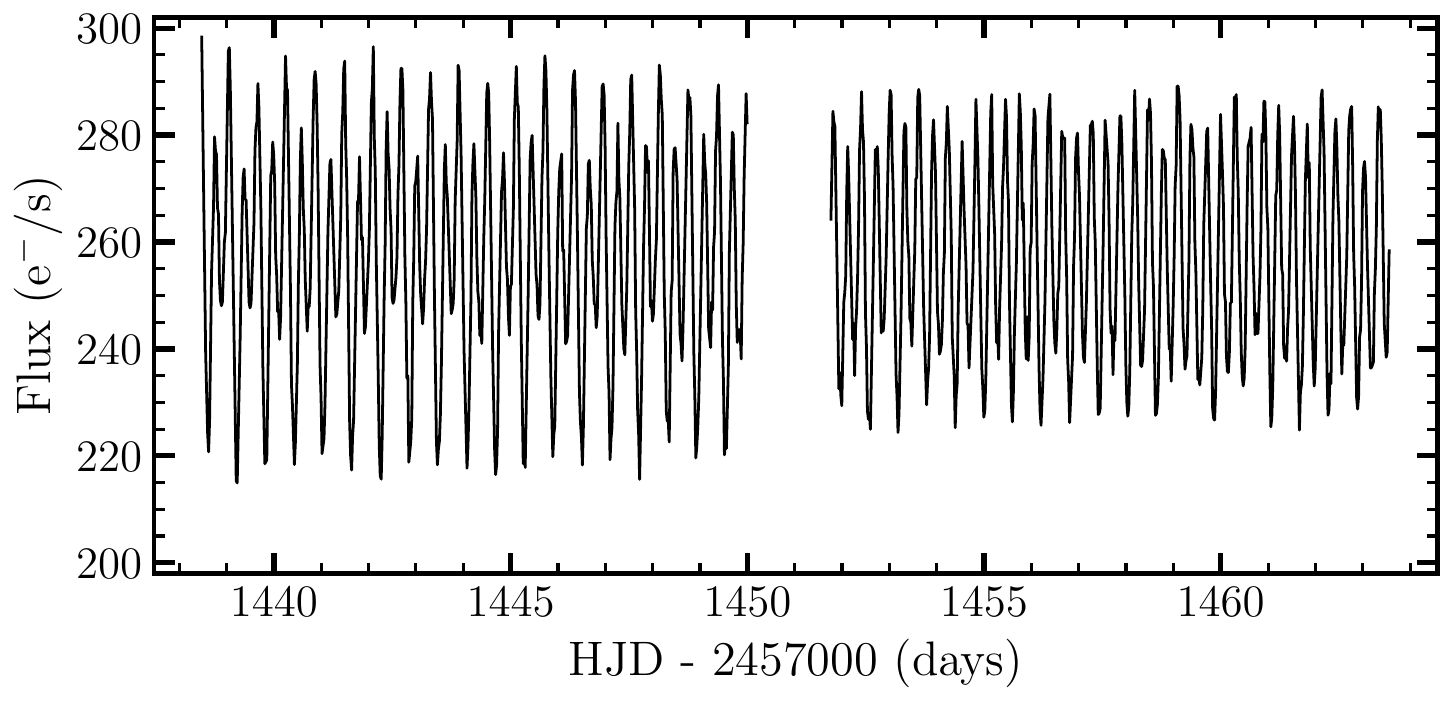}\\
    \includegraphics[width=0.47\textwidth]{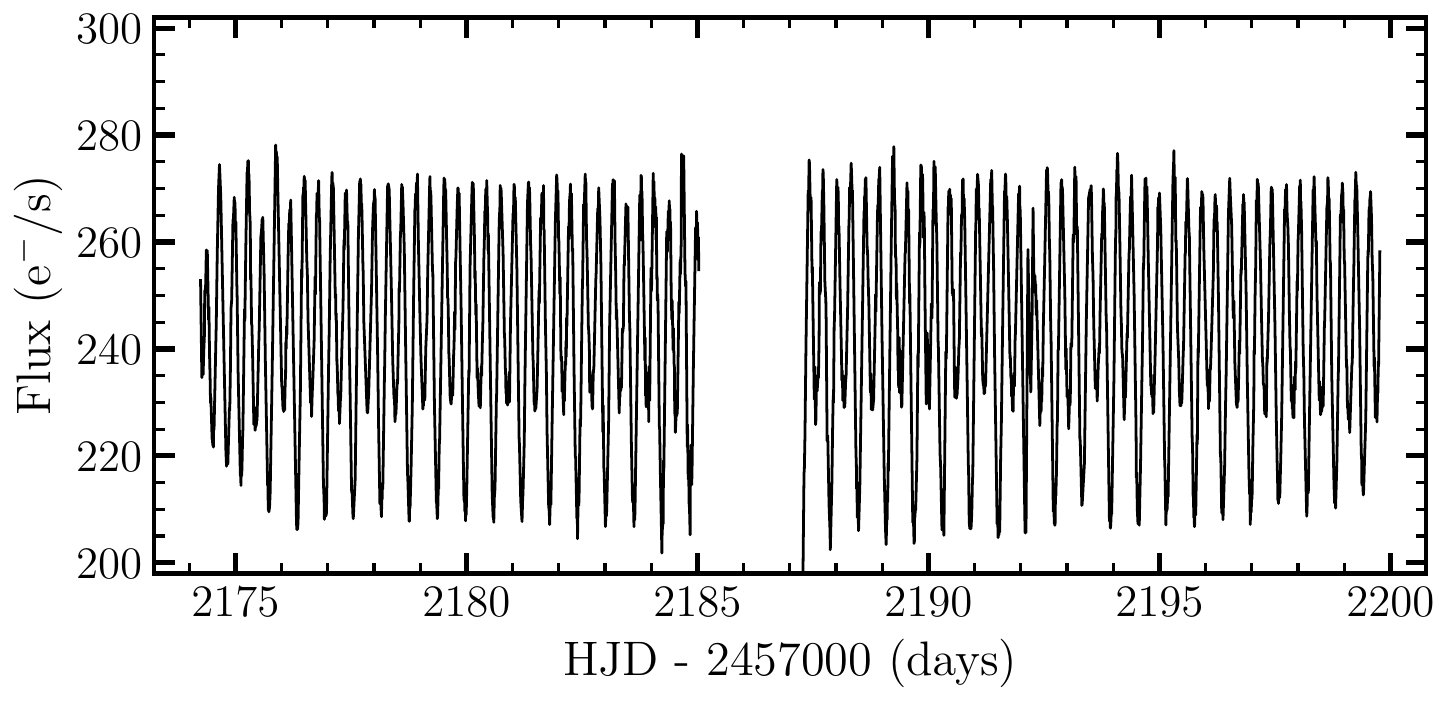}
    \caption{The TESS light curves of J0419. The top panel was observed 
    from 2018-11-15 to 2018-12-11, and the bottom panel was observed from 
    2020-11-20 to 2020-12-16. The light curve of the top panel shows obvious 
    evolution of the peaks and valleys with time.
    }
    \label{fig:tess_lc}
\end{figure}

TESS observed J0419 in two sectors in 2018 and 2020, respectively, using 
the FFIs mode. The first sector was taken from 2018-11-15 to 2018-12-11, 
and the second was from 2020-11-20 to 2020-12-16, with each exposure 
time of 1426\,s, and 475\,s, respectively. A total number of 1176 and 
3589 points were obtained in the two sectors. We use a python package 
\texttt{lightkurve}\footnote{\url{https://docs.lightkurve.org/}} 
\citep{Lightkurve2018} to reduce the data and get the TESS light curves 
of J0419. Images with background counts higher than 150 have been 
eliminated before the light curve extraction, because we cannot obtain 
reliable flux from these seriously contaminated images. After the visual 
inspection, we retain 1111 and 3347 points for the first and second 
observations, respectively. We use a Pixel Level Decorrelation 
(PLD, see \citealt{Deming2015,Luger2016,Luger2018}) method to remove 
systematic instrumental trends. The TESS light curves are shown in 
Figure \ref{fig:tess_lc}. 

\subsubsection{ASAS-SN}

The ASAS-SN is an automated
program to survey the entire visible sky every night down to about 18th 
magnitude \citep{Shappee2014,Kochanek2017}.
For J0419, ASAS-SN observed 
the light curve of $V$-band from 2012-02-16 to 2018-11-29 and $G$-band 
from 2017-09-22 to 2021-07-14. The $V$-band light curve contains 1000
points with a typical uncertainty of 0.051~mag, and the $G$-band light curve 
contains 1943 points with a typical uncertainty of 0.047~mag. We only include 
data points with uncertainties less than 0.1 mag, i.e., 895 data points for 
the $V$-band light curve, and 1656 data points for the $G$-band light curve.

During the ASAS-SN observations, the $V$-band light curve shows a long 
trend of flux increasing from 2014 to 2020 with an amplitude of about 
0.2~mag (see Figure \ref{fig:lc_all}). In addition, three points of the
ASAS-SN $G$-band light curve near 2019 in Figure \ref{fig:lc_all} far exceed 
the mean flux range, which raises the suspicion that there is an outburst 
event. However, the ZTF points observed on the same night have normal 
fluxes, 
and there is no sign of outburst of ASAS-SN points observed on adjacent 
nights. We suspect that the three points are outliers that might be caused 
by unknown instrumental or data processing problems.

\subsubsection{CRTS}

J0419 is in the catalog of CRTS with observation 
time from 2005-10-01 to 2013-10-27. In the eight years of monitoring, 
CRTS has obtained 394 points. The CRTS monitoring was about 7 years earlier
than the ASAS-SN sky survey. During the CRTS monitoring, the light curve was 
stable and did not show a long-term trend or short-term outburst.

\subsubsection{ZTF}

We also collected the optical light curve of J0419 from the 
public DR7 of the 
ZTF program. 
The ZTF $g$-band light curve of J0419 has 340 points with a median flux 
uncertainty of 0.013~mag during the observation from 2018-03-27 to 
2021-03-23. 
The $r$-band has 349 points with median flux uncertainty of 0.012~mag 
during the observation from 2018-03-28 to 2021-03-28. The ZTF data almost 
overlaps the $G$-band light curve of ASAS-SN in time coverage but has a 
higher flux precision.

\section{Data analysis}
\label{sec:data_analysis}

\subsection{\gaia information}

The \gaia DR3 ID of J0419 is 3298897073626626048. We collect the 
astrometric information of J0419 from \gaia early data release 3 
(EDR3; see \citealt{gaia2021}), which provide a parallax of 
$\parallax = 1.45 \pm 0.03$~mas with proper motions of
$\mu_\alpha = 2.17 \pm 0.03$~mas\,yr$^{-1}$ and 
$\mu_\delta = -0.68 \pm 0.02$~mas\,yr$^{-1}$. Based on a parallax 
zero-point correction $zpt = -0.043908$~mas from \citet{Lindegren2021}, 
we obtain a distance of J0419 to the Sun, $d = 671.3 \pm 12.5 \pc$.

\begin{deluxetable*}{llrl}
	\tablenum{2}
	\tablecaption{Orbital and stellar parameters of J0419. The astronomical 
    parameters are from \gaia EDR3. The stellar parameters from SED fitting 
    and spectral fitting are all listed in this table.}
	\label{tab:orbpar}
	\tablewidth{0pt}
	\tablehead{
		\colhead{\textbf{Parameter}} & \colhead{\textbf{Unit}} & \colhead{\textbf{Value}} & \colhead{\textbf{Note}}
	}
	\startdata
\multicolumn{4}{l}{\textbf{The astronomical parameters}} \\
R.A. & h:m:s (J2000) & 04:19:20.07 & Right Ascension\\
Decl. & d:m:s (J2000) & +07:25:45.4 & Declination\\
\gaia parallax & mas & $1.45\pm 0.03$ & The parallax measured by \gaia EDR3\\
$d (\gaia)$ & pc & $671.3 \pm 12.5$ & Distance derived from \gaia EDR3\\
$\mu_\alpha$ & mas\,yr$^{-1}$ & $2.17 \pm 0.03$ & Proper motion in right ascension direction\\
$\mu_\delta$ & mas\,yr$^{-1}$ & $-0.68 \pm 0.02$ & Proper motion in declination direction\\
$G$-band magnitude & mag & $14.70\pm 0.01$ & The $G$-band magnitude measured by \gaia EDR3\\
\hline
\multicolumn{4}{l}{\textbf{The Orbital parameters}} \\
$P_\mathrm{orb}$ & days & 0.6071890(3) & Orbital period\\
$T_0$ & HJD & 2453644.8439(5) & Ephemeris zero-point\\
$K_1$ & \kmps & $216\pm3$ & RV semi-amplitude of the visible star\\
$\gamma$ & \kmps & $86\pm3$ & The systemic RV of J0419 \\
$f(M_2)$ & $M_\odot$ & $0.63 \pm 0.03$ & Mass function of the compact star\\
\hline
\multicolumn{4}{l}{\textbf{Parameters of the pre-ELM}} \\
$T_\mathrm{eff}$ & K & $5793_{-133}^{+124}$ & Effective temperature derived from SED fitting \\
$T_\mathrm{eff}$ (spectral fit) & K & $5776\pm168$ & Effective temperature derived from spectral fitting\\
$\log g$ (spectral fit) & dex & $3.95\pm0.45$ & Surface gravity from spectral fitting\\
$\log g$ & dex & $3.90\pm0.01$ & Surface gravity from SED fitting\\
Metallicity & $\mathrm{[M/H]}$ & $-0.86\pm0.24$ & Metallicity from spectral fitting\\
$M_1$ & $M_\odot$ & $0.176\pm 0.014$ & Mass of the visible star\\
$R_1$ & $R_\odot$ & $0.782_{-0.019}^{+0.021}$ & Effective radius of the visible star \\
$L_\mathrm{bol}$ & $L_\odot$ & $0.62_{-0.10}^{+0.11}$ & Bolometric luminosity of the visible star\\
$A(V)_{\rm SED}$ & mag & $0.34_{-0.10}^{+0.07}$ & The extinction value obtained from the SED fitting\\
\enddata
\end{deluxetable*}

\subsection{Orbital period}
\label{sec:orb_period}

We use the Lomb–Scargle periodogram \citep{Lomb1976,Scargle1982} to 
determine the photometric period of J0419. To improve the accuracy of the 
period value, we use an as-long-as-possible time series from 2005 to 2021 to 
calculate the Lomb–Scargle power spectrum. 
We reject the ASAS-SN data for the concerns that the long trend might 
interfere with the measurement results. The light curves of CRTS, TESS, and 
ZTF are combined after the flux normalization. We estimate the uncertainty 
of the period by using a bootstrap method \citep{Efron1979} that we repeat 
10000 times measurements with randomly removing partial points in each 
measurement. The Lomb–Scargle periodogram gives a period with an error of 
$P_\mathrm{orb} = 0.6071890(3)$ days.
Note that for the ellipsoidal variation, the real orbital period is twice 
the peak period on the Lomb–Scargle power spectrum. 

In order to determine the zero point of ephemeris $T_0$, we use a three-term 
Fourier model \citep{Morris1993},
\begin{equation}
    \begin{aligned}
        f(t) =\ & a_0 \cos [\omega (t-T_0)] + a_1 \cos[2\omega (t-T_0)]\\ 
               & + a_2 \cos [3\omega (t-T_0)],
    \end{aligned}
\end{equation}
to fit the normalized light curve, where $\omega = 2 \pi/ P_\mathrm{orb}$,
$a_0$, $a_1$, $a_2$ are the parameters used to fit the light curve profile. 
We find the best-fitting parameters by minimizing the $\chi^2$ statistics,
which yield the zero point of ephemeris of $T_0 = 2453644.8439(5)$, where 
$T_0$ corresponds to the superior conjunction. We list $P_\mathrm{orb}$ 
and $T_0$ in Table \ref{tab:orbpar}. The folded light curves from different 
surveys or filters using $P_\mathrm{orb}$ and $T_0$ are shown in Figure 
\ref{fig:folded_lc}.

\subsection{Photometric variability}
\label{sec:phot_var}

\begin{figure*}
    \centering
    \includegraphics[width=0.7\textwidth]{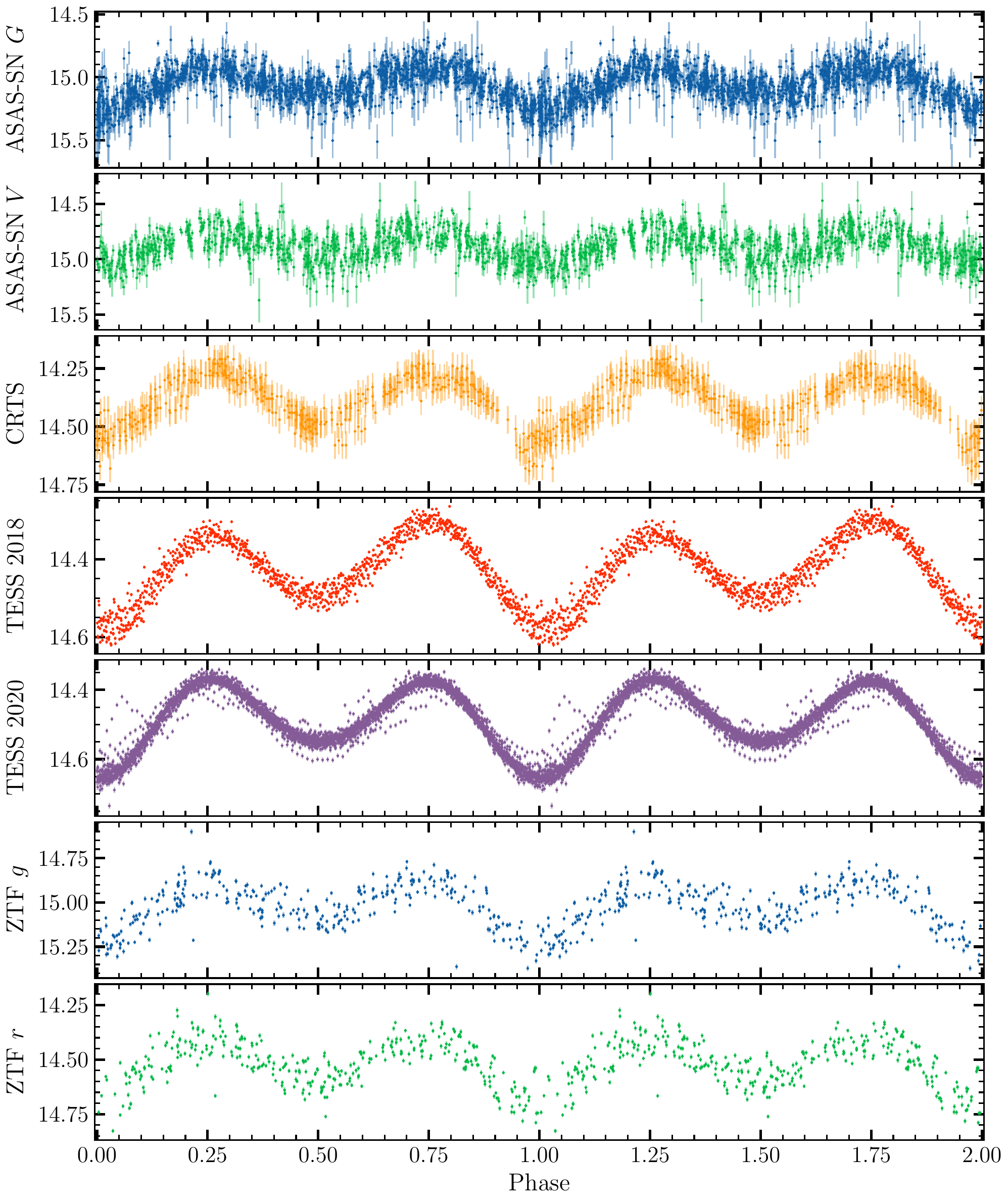}
    \caption{The folded light curves of J0419. The light curves show 
    ellipsoidal variability with a full amplitude of about 0.3~mag. 
    The two sectors of TESS data are shown in two 
    panels, respectively.}
    \label{fig:folded_lc}
\end{figure*}

The folded light curves (Figure \ref{fig:folded_lc}) show ellipsoidal 
variability with amplitudes of about 0.3~mag, together with the evidence of 
mass transfer (the obvious 
emission lines in the spectra), indicating that the visible star is already 
full of the Roche lobe. We did not find any outburst event of this source in 
the 15 years of photometric monitoring, suggesting that the mass transfer 
rate is very low. The result is similar to \citet{elbadry2021b,Badry2021} and different to normal CVs. 

The high cadence TESS observation can be used to show the light curve 
profile at each period. The light curve observed in 2020 (bottom panel 
of Figure \ref{fig:tess_lc}) exhibits a typical ellipsoidal variation, 
while the light curve observed in 2018 (top panel of Figure 
\ref{fig:tess_lc}) has a long-term evolution trend of the peaks and 
valleys beyond the period. The timescale of the evolution seems to be 
several ten days, which is consistent with the timescale of spot activity 
\citep{Hussain2002,Reinhold2013}. We suspect that the long-term 
evolution of the light curve observed in 2018 may result from the spot 
activity.

The folded light curves (except for the TESS data observed in 2020)
show larger scatters than the measurement errors (see the ZTF light 
curves in Figure \ref{fig:folded_lc}). The extra scatter could be due 
to multiple reasons. The spot activity we mentioned above may bring about 
an additional scatter.
The flux from the accretion disk may also contribute to the dispersion of 
the 
light curves, although the mass transfer rate of J0419 is very low. 
The temperature and $\log g$ (see Section \ref{sec:sed_fit}) suggest that 
J0419 falls in the pre-ELM WD instability strip 
\citep{corsico2016,wangkun2020}, in which the pulsation can be driven by the 
$\kappa - \gamma$ mechanism \citep{Unno1989} and the ``convective driving'' 
mechanism \citep{Brickhill1991} acting at the H-ionization and He-ionization 
zones, the scatter may be partly from the pulsation.

\subsection{Radial velocities} \label{sec:rv}

We obtain the template used to measure RV of each single epoch spectrum of 
J0419 through a python package \texttt{PyHammar}\footnote{\url{https://github.com/BU-hammerTeam/PyHammer}}, 
and the best-fitting stellar type is G0. 
The RVs are then measured by using the cross-correlation function (CCF).
The uncertainties of the RVs are estimated using the ``ﬂux 
randomization random subset sampling (FR/RSS)'' method \citep{peterson1998}.
Only the spectral wavelength from 4910\,\AA\ to 5375\,\AA\ is 
used to measure the RVs to avoid the disruption of Balmer and $\rm He~I$ 
emission lines and telluric lines (see Figure \ref{fig:spec_all}). 
Because of the low resolution, the spectra observed by the 2.16-meter telescope 
using the G4 grism are excluded from the RV measurements.

\begin{figure*}
    \centering
    \includegraphics[width=0.9\textwidth]{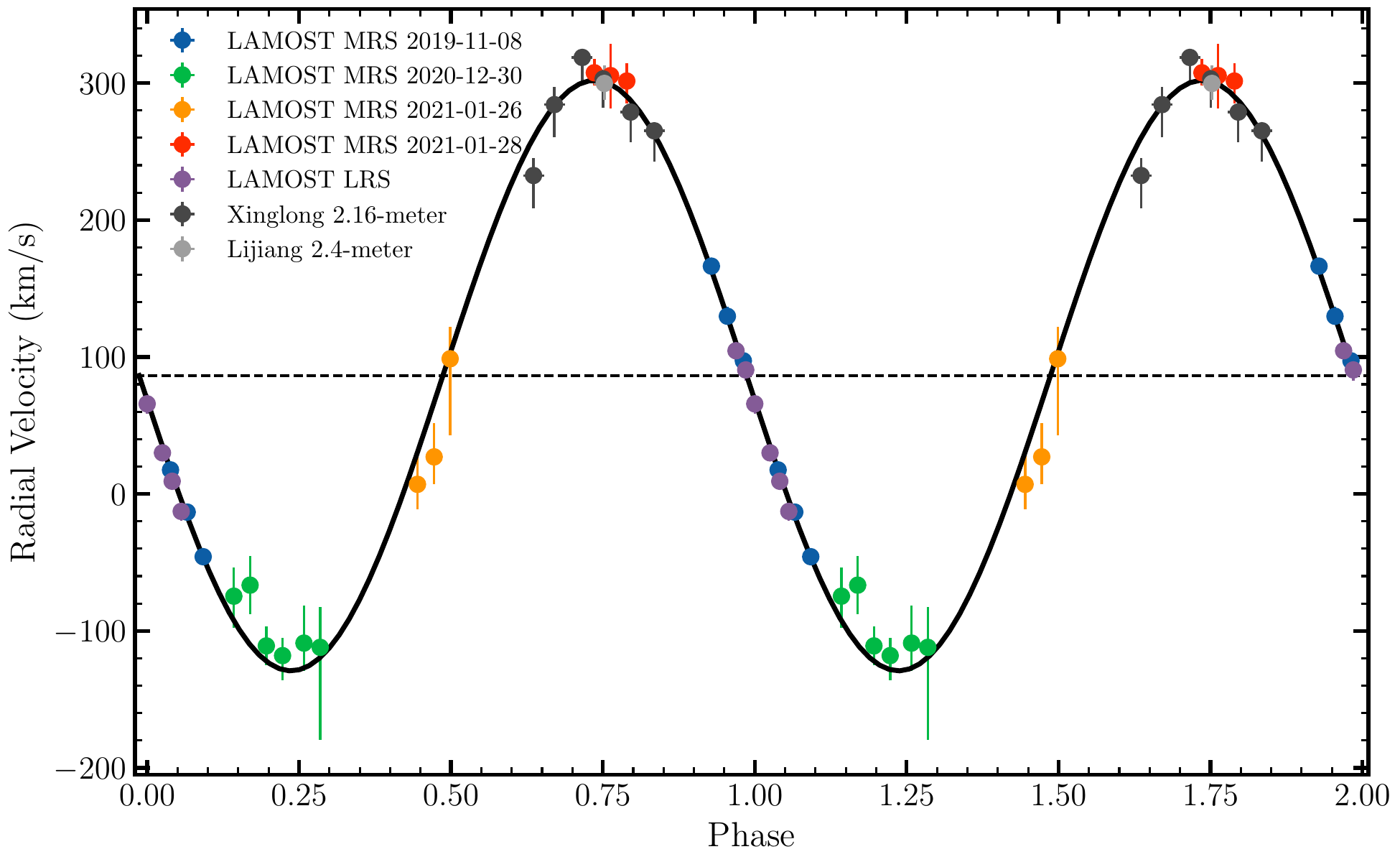}
    \caption{Radial velocities of the visible star. The period used to fold 
    the RVs is $P_\mathrm{orb} = 0.607189$~days. The points observed by 
    different telescopes or nights are plotted with different colors and 
    have been labeled at the top left of the panel. A sinusoidal function is 
    used to fit the RVs. The dashed line represents the systemic RV of 
    $\rm \gamma = 86$\,\kmps, 
    and the black solid line is the best-fit RV 
    curve with a semi-amplitude of 
    $K_1 = 216$\,\kmps}.
    \label{fig:RV_all}
\end{figure*}

We fold the RVs in one phase using the period of 
$P_\mathrm{orb} = 0.607189$~days and $T_0 = 2453644.8439$ derived from 
Section \ref{sec:orb_period} and display the result in Figure 
\ref{fig:RV_all}. Since the visible star is full of Roche lobe, the orbital 
circularization is effective, i.e., the binary moves along a 
circular orbit \citep{Zahn1977}. Therefore, we fit the RVs with a circular 
orbit model following the equation:
\begin{equation}
    V(t) = -K_1 \sin\left[\omega (t + \Delta t)\right] + \gamma,
\end{equation}
where $K_1$ is the semi-amplitude of RVs of the visible star, 
$\omega = 2\pi/P_\mathrm{orb}$, $\gamma$ is the systemic velocity of 
J0419 to the Sun, 
and $\Delta t$ represents the
possible zero-point shift caused by the limited period accuracy when 
folding the RVs.
The fitting results are
$K_1 = 216\pm 3$\,\kmps, 
$\gamma = 86 \pm 3$\,\kmps, and $\Delta t = 12 \pm 3$~minutes.
We display the RV model curve in Figure \ref{fig:RV_all}. 
The best-fitting RV model well matches the measured RVs.

The mass function of a binary is defined as
\begin{equation}
    f(M_2) = \frac{M_2^3 \sin^3 i}{(M_1 + M_2)^2} = \frac{K_1^3 P_{\rm orb}}{2\pi G},
    \label{eq:fm}
\end{equation}
where $M_1$ and $M_2$ are the mass of the visible star and the compact star, 
respectively, $K_1$ is the semi-amplitude of the RVs of the visible star, 
$P_\mathrm{orb}$ is the orbital period, and $i$ is the inclination angle of 
the binary to observer. The mass 
function gives the minimum possible mass of the compact star. 
Using $P_\mathrm{orb} = 0.6071890(3)$~days and 
$K_1 = 216\pm 3$\,\kmps,
we get the mass function of the compact star, 
$f(M_2) = 0.63\pm0.03\,M_\odot$.

\subsection{Spectroscopic stellar parameters}\label{sec:spec_fit}

Our spectra were observed in different telescopes or instruments with very
different wavelength coverage and resolution. It is difficult to generate
a mean spectrum by including all the spectra. Another problem is that most 
of the spectra have obvious emission lines that may interfere with the 
measurements of stellar parameters. Therefore, we only use the 
Lijiang spectrum to take the measurement, which has a good SNR (40.5) 
and weakest emission lines.

\begin{figure*}
    \centering
    \includegraphics[width=0.8\textwidth]{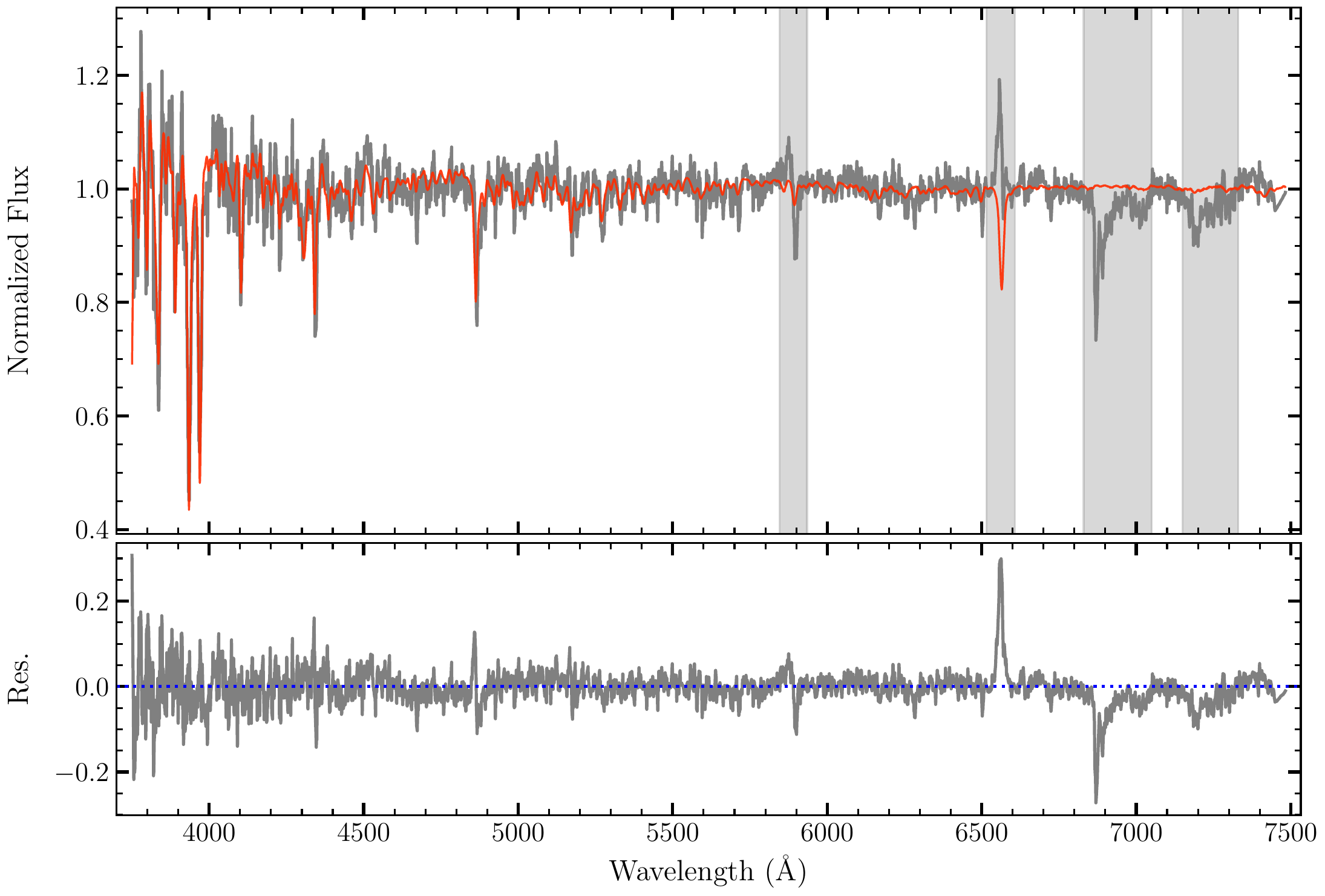}
    \caption{The stellar spectrum fitting result. The gray component in 
    upper panel is the observed spectrum. The red spectrum in upper 
    panel is the model spectrum. The gray component in bottom panel is 
    the residual spectrum. The wavelength of the observed spectrum 
    contaminated by emission lines or telluric lines has been masked 
    before our fitting and shaded with gray in top panel.}\label{fig:spec_fit}
\end{figure*}

We use a python package \texttt{The\_Payne}\footnote{\url{https://github.com/tingyuansen/The_Payne}}
to interpolate the model spectra. 
\texttt{The\_Payne} is a spectral interpolate tool that enable to return a 
template spectrum when we provide a group of stellar parameters. Based on 
a neural-net and spectral interpolation algorithm \citep{Ting2019}, 
\texttt{The\_Payne} can interpolate the spectral grid with 
flexible labels efficiently.

We adopt the BOSZ grid of Kurucz model spectra \citep{Bohlin2017} with five 
labels ($T_\mathrm{eff}$, $\log g$, $\rm [M/H]$, $\rm [C/M]$, $\rm [\alpha/M]$) 
to train the template model. The stellar label intervals provided by BOSZ 
are $\Delta T_\mathrm{eff} = 250\,\mathrm{K}$, 
$\Delta \log g = 0.5$\,dex, $\Delta \mathrm{[Fe/H] = 0.25}$\,dex. 
Before the training, we have 
reduced the resolution of the template to match the
resolution of the observed spectrum, and the flux of the template spectra 
are also normalized using pseudo-continuums that were generated by 
convolving the template spectra with a gaussian kernel 
($\sigma_\mathrm{width} = 50\,\mathrm{\AA}$).

We construct a likelihood function considering both $\chi^2$ and 
a systematic error,
\begin{equation}
    \begin{aligned}
    \ln &  p(f|{\rm \lambda, pms, \sigma_{sys}}) = \\ 
    & -\frac{1}{2}\sum_{\lambda = \lambda_0}^{\lambda_n}\left[ \frac{(f_\lambda - {\rm model}_\lambda)^2}{s_\lambda^2} + \ln (2\pi s_\lambda^2) \right],
\end{aligned}
\end{equation}
where
\begin{equation}
    s_\lambda^2 = \sigma_\lambda^2 + \sigma_\mathrm{sys}^2,
\end{equation}
$f$ represent the normalized spectrum, $\rm pms$ 
represent the stellar labels of the template spectrum, $\sigma_{\rm sys}$ is 
the systematic error. The posteriors of $\rm pms$ and $\sigma_{\rm sys}$ 
are sampled by the software \texttt{emcee} \citep{Foreman2013} based on the 
Markov chain Monte Carlo (MCMC) method. 

The \texttt{emcee} sampling yield a result of $T_{\rm eff} = 5776$\,K,
$\log g = 3.95$\,dex, $\rm [M/H] = -0.86$\,dex.
Similar to \citet{xiang2019} and \citet{Badry2021}, our fitting also underestimate the 
uncertainties of the stellar parameters with small 
$\Delta T_\mathrm{eff} = 56$K, $\Delta\log g = 0.19$~dex, $\Delta \mathrm{[M/H]} = 0.08$~dex. 
For simplicity, we directly use $3\sigma$ of the posterior sample as our 
fitting uncertainties. The fitting results are listed in Table 
\ref{tab:orbpar}.

Figure \ref{fig:spec_fit} shows the fitting result of the Lijiang spectrum. 
The gray spectrum in the top panel is the observation data, and the red 
spectrum is the best-fitting template. The residual spectrum of 
$f_{\rm obs} - f_{\rm model}$ is shown in bottom panel. The shadow region in 
Figure \ref{fig:spec_fit} is masked when we perform the fit.
We find the model spectrum agrees well with the observed spectrum,
except for $\rm H\beta$, $\rm H\alpha$ and several weak $\rm He~I$
emission lines clearly showing on the residual spectrum. 
These emission lines are common in CV spectra \citep[e.g.][]{Sheets2007}. 
The LAMOST spectra with higher resolution show clear double peak emission 
lines, which suggest that the emission lines should
be produced by the accretion disk rather than the visible star.
The emission lines' equivalent widths (EW) vary greatly in the 
different observations. If the EWs of the emission lines reflect the mass 
transfer rate in the binary, the EW variations indicate that the mass 
transfer process is intermittent. We discuss the emission lines more in 
Section \ref{sec:dis_emission}.

Our spectra show strong sodium ``D'' absorption lines beyond the 
template spectrum at wavelengths of 5890\,\AA\ and 5896\,\AA, which is 
similar 
to the result of \citet{Badry2021}. Sodium is thought to originate from the 
CNO-processing that can only reach the surface of a star after most of its 
envelope has been stripped off. Therefore, sodium enhancement is 
generally 
observed in evolved CV donors. The presence of strong sodium absorption 
lines suggests that the visible star of J0419 is an evolved star and has lost 
most of its hydrogen envelope.

\subsection{The broad-band spectral energy distribution fitting}\label{sec:sed_fit}

\begin{table*}
    \centering
    \tablenum{3}
    \begin{tabular}{ccccccc}
    \toprule
         Survey & Filter & $N_\mathrm{obs}$ & $\lambda_\mathrm{effective}$ & AB mag & Vega mag & $\log \lambda f_\lambda$ \\
          & & & ($\mu m$) & (mag) & (mag) & $\log(\mathrm{erg\ s^{-1}\ cm^{-2}})$ \\ 
         \hline
         \multirow{5}{*}{APASS} & Johnson~$B$ & 10 & 0.435 & & $15.55 \pm 0.07$ & $-10.787 \pm 0.026$ \\
         & SDSS~$g$ & 10 & 0.472 & $15.13 \pm 0.05$ & & $-10.689 \pm 0.020$ \\
         & Johnson~$V$ & 10 & 0.550 & & $14.83 \pm 0.07$ & $-10.633 \pm 0.027$ \\
         & SDSS~$r$ & 10 & 0.619 & $14.61 \pm 0.07$ & & $-10.597 \pm 0.026$ \\
         & SDSS~$i$ & 10 & 0.750 & $14.42 \pm 0.07$ & & $-10.606 \pm 0.028$ \\
         \hline
         \multirow{5}{*}{Pan-STARSS} & PS1~g & 8 & 0.487 & $14.98 \pm 0.03$ & & $-10.643 \pm 0.014$ \\
         & PS1~$r$ & 10 & 0.621 & $14.69 \pm 0.03$ & & $-10.633 \pm 0.012$ \\
         & PS1~$i$ & 22 & 0.754 & $14.49 \pm 0.02$ & & $-10.637 \pm 0.008$ \\
         & PS1~$z$ & 16 & 0.868 & $14.35 \pm 0.02$ & & $-10.641 \pm 0.009$ \\
         & PS1~$y$ & 13 & 0.963 & $14.30 \pm 0.03$ & & $-10.669 \pm 0.010$ \\
         \hline
         \multirow{3}{*}{2MASS} & 2MASS~$J$ & 1 & 1.241 & & $13.43 \pm 0.1$ & $-10.787 \pm 0.040$ \\
         & 2MASS~$H$ & 1 & 1.651 & & $13.10 \pm 0.1$ & $-10.968 \pm 0.040$ \\
         & 2MASS~$K_\mathrm{s}$ & 1 & 2.166 & & $13.03 \pm 0.1$ & $-11.243 \pm 0.040$ \\
         \hline
         WISE & $\mathrm{W_1}$ & 26 & 3.379 & & $12.97 \pm 0.02$ & $-11.745 \pm 0.010$\\
         & $\mathrm{W_2}$ & 26 & 4.629 & & $12.91 \pm 0.03$ & $-12.115 \pm 0.011$\\
         \hline
    \end{tabular}
    \caption{The SED of J0419. The number of 
    observations is shown in the third column, where the APASS data is a 
    combination of DR9 and DR10. Both APASS and Pan-STARRS data have added 
    additional systematic errors caused by sampling. The 
    magnitudes of 2MASS have been increased by 0.1~mag to correct the phase 
    offset, and the errors have been increased to 0.1~mag. The AB and Vega 
    magnitude systems are displayed in two columns.}
    \label{tab:mag}
\end{table*}

We use the broad-band spectral energy distribution (SED) to constrain 
the stellar parameters and the extinction of J0419. In the SED fitting, 
the peak wavelength from UV to optical can be used to constrain the 
effective temperature, 
$T_\mathrm{eff}$. The deviation of the SED slope from Rayleigh-Jeans law in 
the mid-infrared band is generally considered to be caused by extinction, 
which can be used 
to estimate the extinction value \citep{Majewski2011}. The color of 
$u$-band $v.s.$ other-band represents the metal abundance, $\mathrm{[Fe/H]}$ 
 \citep{huang2021}. If we know the distance of a source, we can get the 
effective radius from the SED fitting.

We use a python 
package \texttt{astroARIADNE}\footnote{
\url{https://github.com/jvines/astroARIADNE}} to fit the SED of J0419. 
\texttt{astroARIADNE} is designed to fit broadband photometry automatically 
that based on a list of stellar atmosphere models by 
using the Nested Sampling algorithm. 
The fitting parameters in \texttt{astroARIADNE} 
include $T_\mathrm{eff}$, $\log g$, [Fe/H], distance, 
stellar radius, extinction parameter $A(V)$.  

\begin{figure*}[ht]
    \centering
    \includegraphics[width=0.9\textwidth]{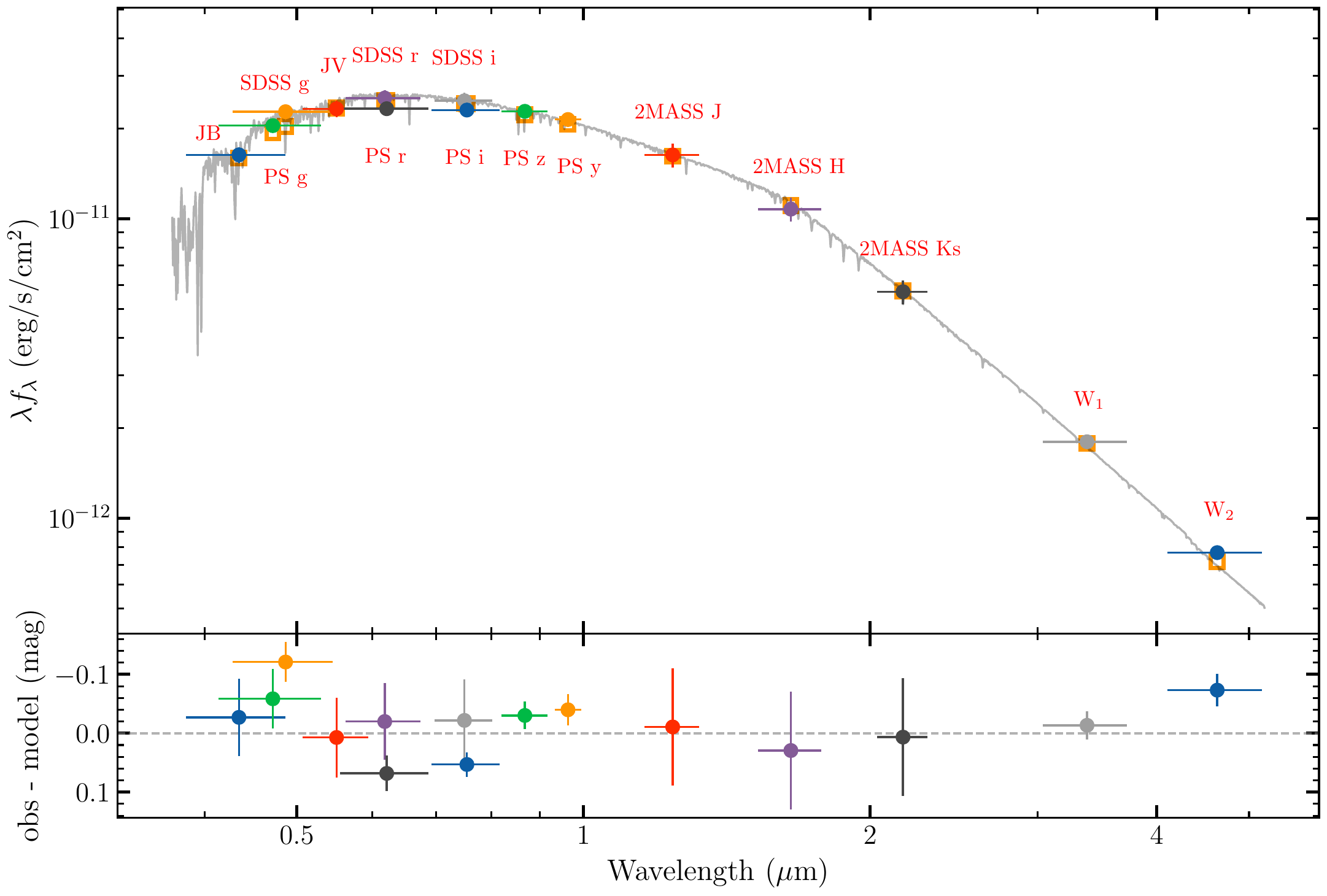}
    \caption{The SED fitting of J0419. The multiband photometric points are 
    plotted in the top panel, and the filter information are displayed near 
    the data points. The model data is also shown in top panel with orange 
    open squares. The model template spectrum is plotted in top panel with 
    gray. The residuals of $f_{\rm obs} - f_{\rm model}$ are plotted in bottom panel.}
    \label{fig:sed}
\end{figure*}

We collect multi-band photometric data of J0419. 
Because GALEX \citep{Martin2005} and $Swift$ \citep{Gehrels2004} did not 
observe J0419, we do not have the UV points. The photometric data flag of 
J0419 in SDSS survey \citep{York2000} suggest that its magnitudes may have 
problems. We therefore use the APASS 
\citep[$B$, $g$, $V$, $r$, and $i$ bands;][]{Henden2015} data instead. 
Our photometric data also include Pan-STARRS 
\citep[$g$, $r$, $i$, $z$, and $y$ bands;][]{Chambers2016}, 
2MASS \citep[$J$, $H$, and $K_\mathrm{s}$ bands;][]{Skrutskie2006}, and 
WISE \citep[$\rm W_1$, $\rm W_2$ bands;][]{Wright2010}. 

For the APASS survey, the DR9 data includes six detections, and DR10 includes 
four detections. We merge these two datasets using the inverse of the error 
as the weight. For the Pan-STARRS survey, the numbers of single epoch 
detections in each band are 8, 10, 22, 16, and 13, respectively. 
Considering that J0419 shows significant variations, we add
systematic uncertainties caused by random sampling of the light curves 
to the errors of the photometric data. The systematic uncertainties in each band 
are estimated as $\sigma_\mathrm{sys} = \mathrm{std} / \sqrt{N}$, where 
$\mathrm{std}$ is the mean standard deviation of the light curves of J0419, 
$N$ is the number of observations in a band. 
2MASS survey only observed J0419 once on 1999-12-07, 08:27:34.51, and the 
corresponding phase is 0.28. We assume that the IR band light curve 
has the same variation amplitude as the 
optical band to calculate the deviation between the observed magnitude 
and the mean magnitude and add 0.1~mag to 2MASS data to correct the 
deviation. Considering the time interval between 2MASS and other surveys 
(10 to 20 years) and possible light curve trend, 
we increase the magnitude uncertainties of 2MASS to 0.1~mag.
All the photometric data are summarized in Table \ref{tab:mag}.

The visible star of J0419 has filled the Roche lobe 
(Section \ref{sec:phot_var}). In this case, the mean density is given by
\begin{equation}
    \bar{\rho} = \frac{3 M_1}{4 \pi R_1^3} \cong 110 P_\mathrm{hr}^{-2}\,\mathrm{g\,cm^{-3}},
    \label{eq:rourobe}
\end{equation}
where $M_1$ is the mass, $R_1$ is the equivalent radius of the 
Roche-filling star, the period, $P_\mathrm{hr}$, is in the unit of hours 
\citep{Frank2002}. Equation \ref{eq:rourobe} shows that 
the mean density of Roche lobe only depends on the orbital period.
Combining the radius derived from \texttt{astroARIADNE} fitting and the 
mean density of Roche lobe, we can calculate the mass and \logg of the 
visible star. 

We fit the SED in the following way. First, we use the distance measured by 
\gaia EDR3 as the prior of distance parameter, and other parameters are 
set to default values. Then we use the radius from the fitting result to 
calculate the mass and
$\log g$ of the visible star (Equation \ref{eq:rourobe}).
Second, we update the \logg prior by using our calculation and re-fit the 
SED. 
\logg and $\rm [Fe/H]$ only have little effect on the SED, so the 
fitting parameters converge quickly.
The radius obtained from the SED fitting is 
$R_1 = 0.782_{-0.019}^{+0.021} \,R_\odot$, and the corresponding visible 
star mass is $M_1 = 0.176\pm0.014\, M_\odot$. The effective 
temperature of the visible star is 
$T_\mathrm{eff} = 5793_{-133}^{+124}\, \mathrm{K}$. The bolometric 
luminosity derived from the SED fitting is 
$L_\mathrm{bol} = 4\pi R_1^2\,\sigma T_\mathrm{eff}^4 = 0.62_{-0.10}^{+0.11}\, L_\odot$. 
We summarize the SED fitting result in Table \ref{tab:orbpar}, and show 
the best fit model in Figure \ref{fig:sed}. 

According to the mass function, 
$f(M_2) = 0.63\,M_\odot$
and the 
visible star mass $M_1 = 0.176 \,M_\odot$, we obtain the minimum 
mass of the compact star, 
$M_2 \geq 0.9\,M_\odot$. 
If we assume the compact 
star is a WD and estimate its radius using the mass-radius relation of 
\citet{bedard2020}, the corresponding WD radius is 
$R_\mathrm{2,max} < 0.01 R_\odot$. Even if the temperature of the compact 
star is 20,000\,K, its flux contribution in total luminosity is less than 
3\% and is negligible. The flux contribution from the accretion disk is 
more complex and will be discussed in Section \ref{sec:dis_poscontrib}. 
Our analysis in that Section demonstrate that the flux 
contribution in optical band is dominated by the visible star. 

\subsection{The light curve fitting}

\begin{figure*}
    \centering
    \includegraphics[width=0.8\textwidth]{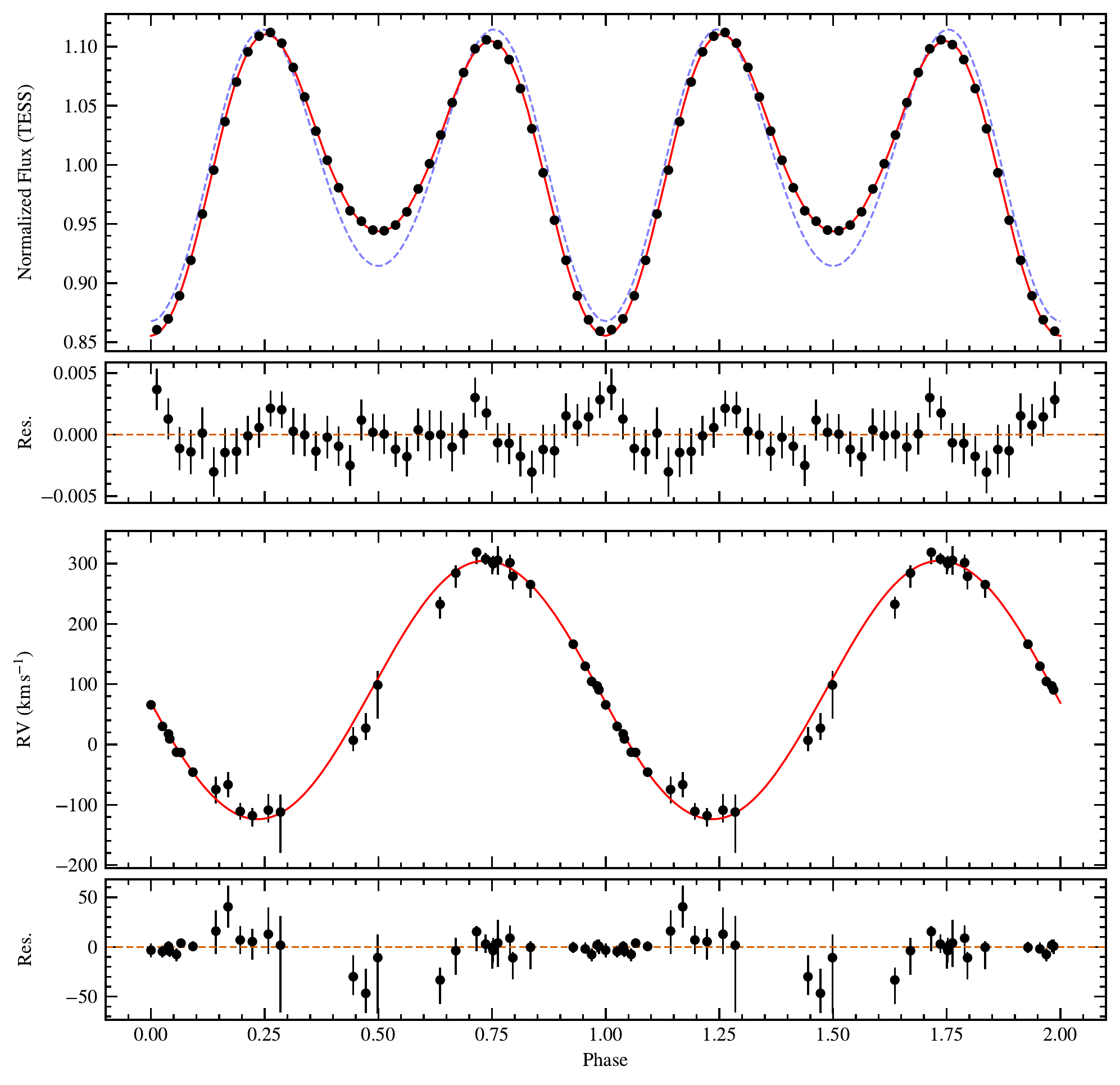}
    \caption{
        Best-fit light and RV curves. The top panel shows the 
        rebinned TESS light curve (filled circles). The dashed blue curve 
        represents the best-fitting ellipsoidal variability model, which 
        cannot well match the observations. Hence, we add a star spot to 
        the model and refit the TESS light curve, and the best-fitting 
        result is shown as the solid red curve. Indeed, adding a spot
        improves the fit result 
        greatly. The second panel indicates the residuals between the 
        observed and model fluxes. Our model with a star spot well matches 
        the observed data, 
        and the reduced $\chi^2$ is 1.1 (close to 1.0). The third panel 
        displays the 
        RV curve, in which the circles are the observed data, and the 
        solid curve is the best-fitting RV model (with a star spot).
        The bottom panel presents the 
        residuals between the observed RV curve and the model RV curve.}
    \label{fig:lcfit}
\end{figure*}

\begin{figure*}
    \centering
    \includegraphics[width=\textwidth]{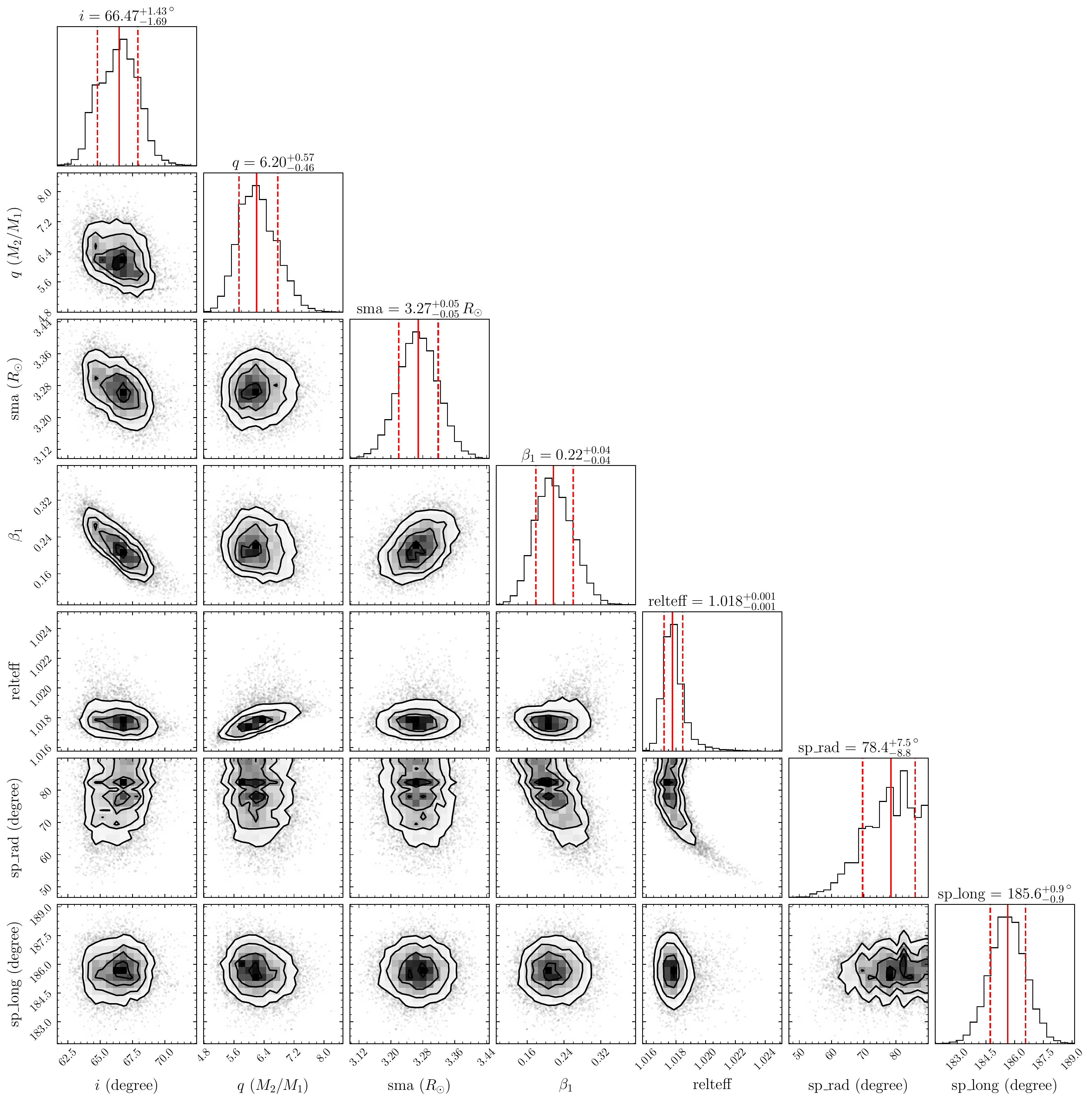}
    \caption{
        Parameter distributions from the joint fitting of light and RV 
        curves. We only illustrate a fraction of parameters, and other 
        parameters are summarized in Table \ref{tab:lc_fit}.}
    \label{fig:lcfit_corner}
\end{figure*}

J0419 exhibits ellipsoidal variability. The main parameters of modulating 
the ellipsoidal light curves are the inclination angle $i$, the mass 
ratio $q = M_2 / M_1$, the filling factor $f_\mathrm{fill}$, the limb 
darkening factors, and the gravity darkening exponent $\beta_1$ 
\citep{von1924}. To estimate the inclination of the binary, we use 
\texttt{Phoebe 2.3}\footnote{\url{http://phoebe-project.org/}} 
\citep{prsa2005,prsa2016,Conroy2020} to model the light curve and RV curve 
of J0419, and inversely solve the orbital parameters.
\texttt{Phoebe} is an open-source software package, which is based on the 
Wilson-Devinney software \citep{wd1971}, for computing light and 
RV curves of binaries using a superior surface discretization algorithm.
Main physical effects in a binary system have been considered, including 
eclipse, the distortion of star shape due to the Roche potential, radiative 
properties (atmosphere intensities, gravity darkening, limb darkening, 
mutual irradiation), and spots.

In the light curve fitting, we set the temperature of the donor to 5793\,K 
derived from the SED fitting (Section \ref{sec:sed_fit}) with 
the \texttt{Phoenix} atmosphere model.
We adopt the logarithmic limb darkening law and obtain the coefficient 
values self-consistently from the \texttt{Phoebe} atmosphere model. 
The gravity darkening law states that, 
$T_\mathrm{eff}^4 \propto g_\mathrm{eff}^{\beta_1}$, where $\beta_1$ is 
the gravity darkening exponent. It is generally assumed that, for 
stars in hydrostatic and radiative equilibrium 
($T_\mathrm{eff} \gtrsim 8000$\,K), $\beta_1 = 1$ \citep{von1924}, and 
for stars with convective envelopes ($T_\mathrm{eff} \lesssim 6300$\,K), 
$\beta_1 = 0.32$ \citep{Lucy1967}. The theoretical dependence of 
$\beta_1$ upon $T_{\mathrm{eff}}$ is obtained by 
\citet[][see their Figure 2]{Claret2011}. J0419 appears to be in the 
temperature range where the transition from convection to equilibrium 
occurs, therefore we set $\beta_1$ as a free parameter and adopt a 
common normal distribution, $\beta_1 \sim \mathcal{N}(0.32, 0.1)$, 
as its prior, just similar to \citet{Badry2021}.

Because the visible star of J0419 is filling the Roche lobe, we set the 
model to be semi-detached ($f_\mathrm{fill} = 1$). We use an equivalent 
radius of $R_1 \sim \mathcal{N}(0.78, 0.02)\,R_\odot$ obtained from the 
SED fitting as the prior of the radius parameter. The free parameters 
in our fit are $i$, $q$, $\beta_1$, $\gamma$, and sma, where sma is the semi-major axis of a binary orbit.

Except for the second TESS observation data, the folded light curves show 
larger scatters than the measurement errors. We, therefore, only fit the 
second TESS data. To reduce the calculation efforts of the model, we 
rebin the light curve of TESS to 40 points. The errors of the rebinned 
light curve include both the measurement errors and a systematic error 
that is estimated using a median filter method \citep{zhang2019}.

We find that a pure ellipsoidal model cannot well explain the observed data. 
Indeed, the residuals between the observed and model fluxes depend upon 
the phases (see Figure \ref{fig:lcfit}). Thus, 
we add a spot to the model to compensate for the phase-dependent residuals.
The spot component is defined with four parameters, relteff, radius, 
colatitude, and longitude. 
The relteff parameter is the ratio of the spot temperature to
the local intrinsic temperature of the star. The radius parameter represents
the spot angular radius. 
The remaining two parameters, colatitude and longitude, indicate the
colatitude and longitude of the spot on the stellar surface, respectively.
We only set relteff, radius, and longitude to be free parameters.
The colatitude is fixed to be 90 degrees.
We use the Akaike Information Criterion (AIC) and Bayesian Information 
Criterion (BIC) to compare the pure ellipsoidal model with the one 
with a spot. For the model without spot, the AIC and BIC of the best 
fitting result are -105 and -85, respectively. For the model with a spot, 
the AIC and BIC of the best fitting result are -282 and -255, 
respectively. Hence, adding a spot improves the fitting result greatly. 
Figure \ref{fig:lcfit} illustrates the best fit model with a spot, 
which is in good agreement with the 
observed light curve. The residuals between the TESS light curve and the 
model with a spot are much smaller than the variability amplitude. 
The light curve residuals in Figure \ref{fig:lcfit} show a weak 
periodic structure. Considering that 
the reduced $\chi^2$ is 1.1 (close to 1.0), the above structure 
is statistically insignificant.
The fitting yields an inclination angle of 
$i = {66.5}_{-1.7}^{+1.4}$ degrees and a mass of the compact object of 
$M_2 = {1.09}_{-0.05}^{+0.05}\,M_\odot$. Figure \ref{fig:lcfit_corner} 
illustrates the distributions of the model parameters  (see also Table \ref{tab:lc_fit}).
We stress that the inferred inclination angle would 
not significantly change if we omit the spot component.

\begin{deluxetable}{lrlr}
    \tablenum{4}
    \tablecaption{Parameters from the joint fit of light and RV curves.}
    \label{tab:lc_fit}
    \tablewidth{0pt}
    \tablehead{
    \colhead{Parameter} &
    \colhead{Value} & \colhead{Parameter} & \colhead{Value}
    }
    \startdata
    $i$ (deg) & ${66.5}_{-1.7}^{+1.4}$ & $q$ ($M_2 / M_1$) & ${6.2}_{-0.5}^{+0.6}$ \\
    sma ($R_\odot$) & ${3.27}_{-0.05}^{+0.05}$ & $\gamma$ (\kmps) & ${87.5}_{-3.6}^{+3.6}$ \\
    $\beta_1$ & ${0.22}_{-0.04}^{+0.04}$ & spot relteff &  ${1.018}_{-0.001}^{+0.001}$ \\
    spot radius (deg) & ${78}_{-9}^{+8}$ & spot long (deg) & ${186}_{-1}^{+1}$ \\
    t0\_rv (minutes) & ${11.7}_{-2.8}^{+2.7}$ & $R_1\,(R_\odot)$ & ${0.771}_{-0.022}^{+0.020}$ \\
    $M_1\,(M_\odot)$ & ${0.177}_{-0.015}^{+0.014}$ & $M_2\,(M_\odot)$ & ${1.094}_{-0.049}^{+0.053}$ \\ 
    $K_1$ (\kmps) & ${214.8}_{-3.4}^{+3.4}$ & & \\
    \enddata
    \tablecomments{
        The spot long is the longitude of the spot,
        where 0 means that the spot points towards the companion of the 
        binary. 
        The t0\_rv parameter accounts for the small phase offset.}
    \end{deluxetable}

\section{Discussion}
\label{sec:discussion}

\subsection{The properties of the emission lines}
\label{sec:dis_emission}

J0419 has nine nights of observations, and most of the nights have taken 
multiple spectra. We normalize each single exposure spectrum and measure 
the EW of $\rm H\alpha$ emission line after subtracting 
the continuum component, where we generate the continuum template by 
using \texttt{The\_Payne} with the stellar parameters obtained in Section 
\ref{sec:spec_fit}.
We obtain the EW of the \ha\ emission line by integrating each residual 
spectrum from 6520\,\AA\ to 6610\,\AA. The EWs of \ha\ are listed in 
Table \ref{tab:spec_stat}. As can be seen from Figure \ref{fig:spec_all}, 
the EWs of the \ha\ emission line have changed greatly from night to night. 
But at the same night, or two nearby observation nights, the EWs change 
little. These show that the timescale of the variations of emission lines 
is from several days to tens of days. 
Some works found the flickering timescale of the emission lines is 
similar to the continuum in CVs \citep[e.g.][]{Ribeiro2009}, ranging from 
minutes to hours. The mechanisms of the flicker could be condensations 
in the matter 
stream \citep{Stockman1979}, non-uniform mass accretion, or turbulence in the 
accretion disk \citep{Elsworth1982}. The longer variability timescale of the 
emission lines of J0419 may be due to the low mass transfer rate.

According to the mass function (Equation \ref{eq:fm}) and visible star mass, 
the mass ratio is $q = M_2 / M_1 > 5.3$. If we assume that the emission 
lines originate 
from the accretion disk, the RV semi-amplitude of the emission lines will be 
$K_\mathrm{em} < 41.3$\,\kmps. The resolution and SNR of the J0419 spectra 
are not enough to measure the RVs of the emission lines. 
However, we find that the wavelength shift of the emission lines is much 
smaller than the continuum component, which disfavors the stellar origin of 
the emission lines.

\begin{figure*}[htp]
    \centering
    \includegraphics[width=0.8\textwidth]{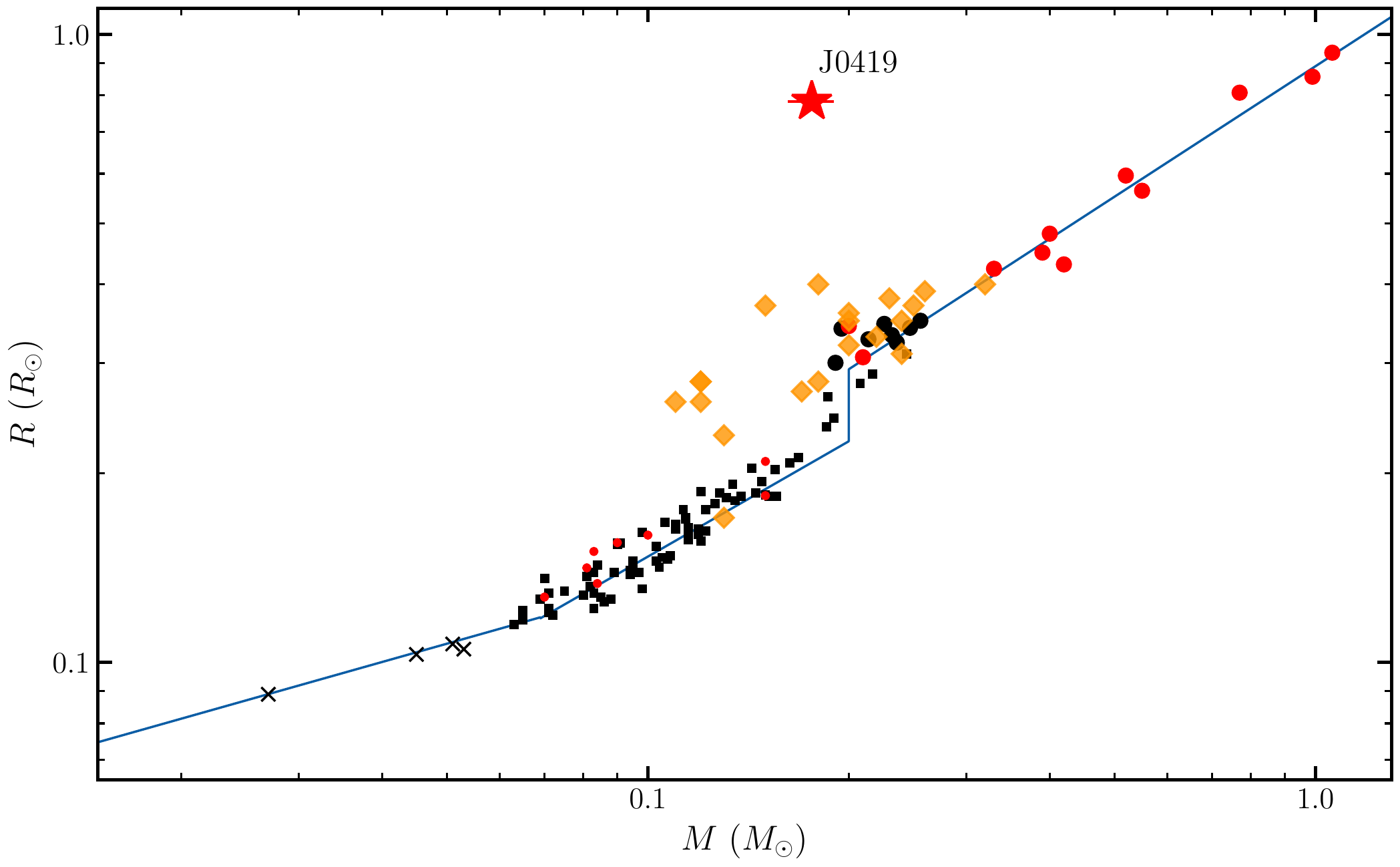}
    \caption{The mass-radius distribution of CVs and pre-ELMs. The red and 
    black points are normal CVs from \citet{Patterson2005}. The orange 
    diamond points are pre-ELMs from \citet{elbadry2021b}, and the red star 
    is J0419. The solid line is the mass-radius relation of CVs 
    adopted from \citet{Knigge2011}.
    }
    \label{fig:m1r1}
\end{figure*}

\subsection{Possible flux from the compact star or disk}
\label{sec:dis_poscontrib}

The radiation from the companion star or disk for an 
accreting binary system will lead to an overestimation of the radius and mass 
of the donor. For J0419, as we mentioned in Section \ref{sec:sed_fit}, due 
to the large mass of the compact star, its flux contribution in total 
luminosity is negligible. The radiation from the accretion disk is more 
complicated. Most normal CVs have strong radiation from disks in the 
optical band. However, for the evolved donors with higher temperatures, 
their mass transfer rate is very low \citep{elbadry2021b,Badry2021} and 
the donors dominate the luminosities.
The spectra of J0419 are 
also clearly dominated by the donor. We list the reasons below:
\begin{enumerate}
    \item[(1)] The SED is well fitted by a pure stellar model; 
    \item[(2)] The template spectrum matches well with the observation 
    spectra; 
    \item[(3)] The light curves of J0419 show ellipsoidal variability, and 
    no CV characteristics were found, such as violent variability in a short 
    timescale, outburst events.
\end{enumerate}

\subsection{Comparison to CVs}

Most CVs have low mass donors whose orbital 
periods are several hours (Section \ref{sec:intro}). The donors show characteristics similar to 
main-sequence stars with the same mass as they have the same chemical 
composition and 
structure. Unlike normal CVs, the pre-ELMs in CV stage have evolved and 
do not follow the donor sequence. Evolved donors generally have higher 
temperatures and possibly more bloated radii than normal CVs with 
similar donor masses.

We collect the mass and radius data of normal CVs from 
\citet{Patterson2005}, and pre-ELM from \citet{elbadry2021b}, and plot them 
in Figure \ref{fig:m1r1}. The red and black points in Figure 
\ref{fig:m1r1} are normal CVs; the orange diamond points are pre-ELMs. 
The solid line is the empirical mass-radius relation of normal CVs from 
\citet{Knigge2011}. J0419 is labeled by the red star for comparison.
We can see that J0419 completely 
deviates from the empirical mass-radius relation of normal CVs. Objects of 
\citet{elbadry2021b} either fall on the mass-radius relation or slightly 
deviate from it, although their temperatures are significantly higher than 
the normal CV donors with the same mass. These show 
that the visible star of J0419 is very bloated and more evolved than the 
objects of \citet{elbadry2021b}.

Compared with normal CVs, the SED of J0419 is dominated by the donor star, 
and emission lines in the spectra are weaker. No outburst events were 
detected in 15 years of monitoring. 
We do not find random variability in a short timescale 
from hours to days in the high cadence TESS light curves, which is 
different from the light curves of most normal CVs \citep{Bruch2021}. 
The objects of \citet{elbadry2021b} exhibit similar properties. 
These indicate that the mass transfer rate of pre-ELM is very low compared 
with normal CVs.

\subsection{Comparison to other pre-ELMs} \label{sec:cmp_pre_ELMs}

\begin{figure}
    \centering
    \includegraphics[width=0.48\textwidth]{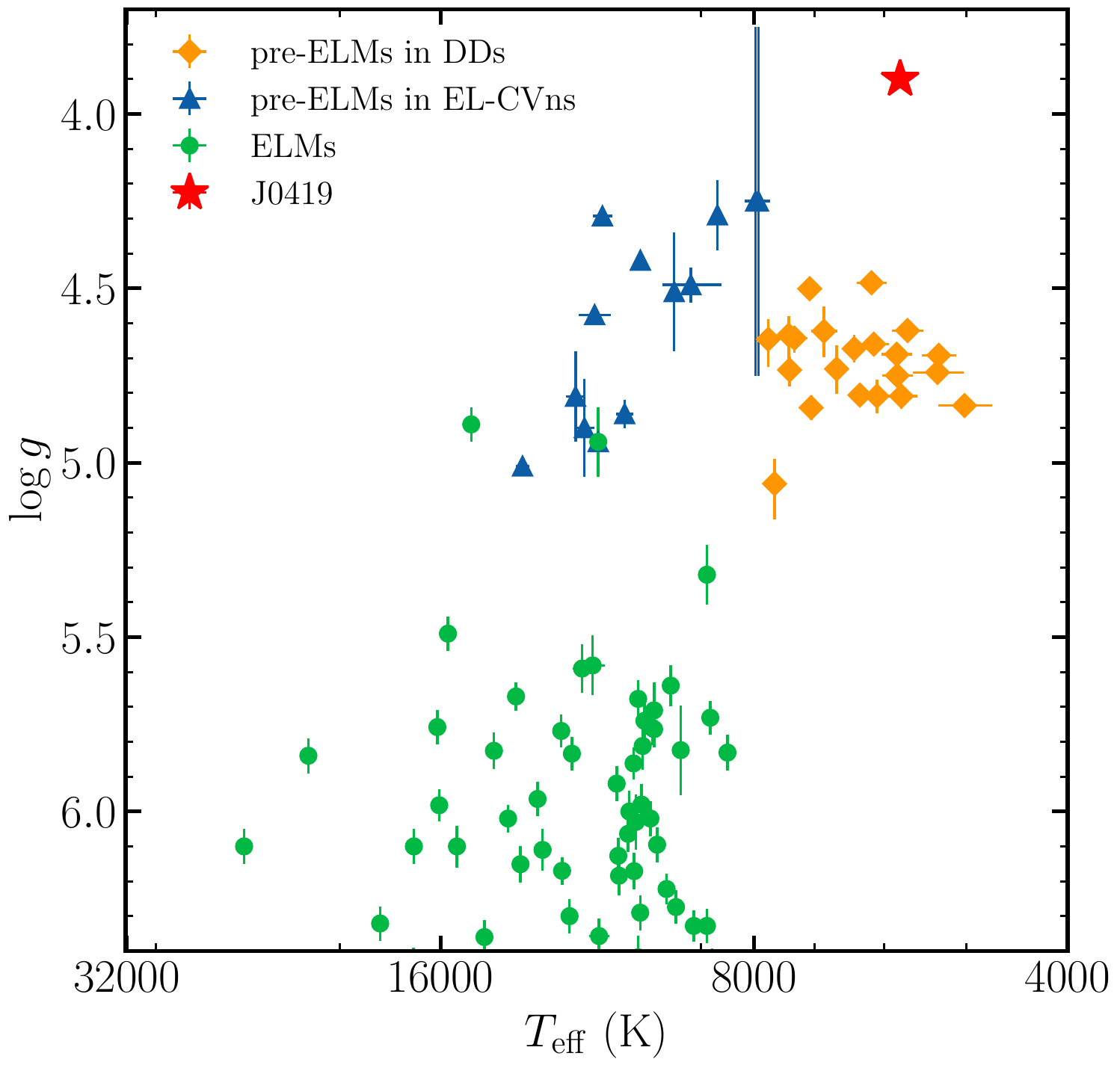}
    \caption{ $T_\mathrm{eff}-\log g$ diagram of ELMs and pre-ELMs. The 
    meaning of different points is labeled in top left of the panel.}
    \label{fig:teff_logg}
\end{figure}

\begin{figure*}
    \centering
    \includegraphics[width=0.9\textwidth]{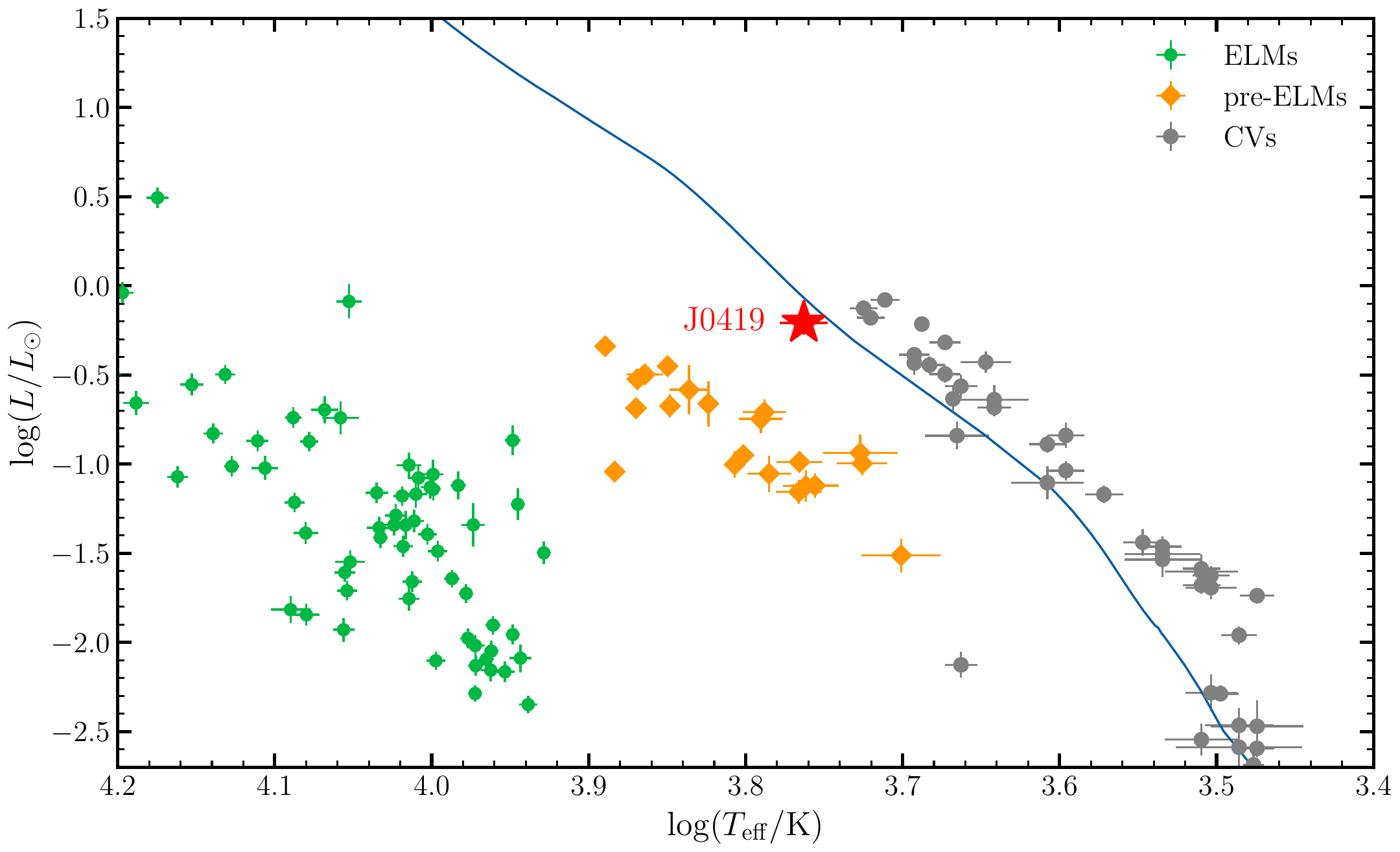}
    \caption{HR diagram. The green points are the ELMs from 
    \citet{brown2020}. The orange diamond points are pre-ELMs from 
    \citet{elbadry2021b}. The gray points are CVs from \citet{Knigge2006}.
    The solid line is the main sequence obtained from \texttt{isochrones} 
    \citep{isochrones2015}. The red star is J0419.}
    \label{fig:HR}
\end{figure*}

Several works have reported the pre-ELMs. These objects are stripped stars 
with burning hydrogen envelopes that more bloated and cooler than ELMs. Most 
of the reported pre-ELMs are found in EL CVns. Their companions are 
main-sequence A or F stars. 
We collect ELMs/pre-ELMs to compare their properties. 
The EL CVn-type stars are from \citet{Maxted2013,Maxted2014a,corsico2016,Corti2016,Gianninas2016,zhang2017,wangkun2020,lee2020}, the pre-ELMs in DDs 
are from \citet{elbadry2021b}, and the ELMs are from \citet{brown2020}.

The pre-ELMs in EL-CVns are hotter than pre-ELMs in DDs (see Figure 
\ref{fig:teff_logg}), which should be due to the selection effect. 
Most of the reported EL CVns are detached binaries 
selected from the eclipse systems. 
Their radii have been shrinking after the detachment, accompanied by the 
increase of surface temperatures.
The reported pre-ELMs (including 
J0419 and the sources of \citealt{elbadry2021b}) in DDs all 
have distinct ellipsoidal variability. They are filling or close filling 
the Roche lobe with lower temperatures. Hence, the pre-ELMs in EL-CVns are 
at the latter stage of the evolution than the reported pre-ELMs in DDs.

For J0419, its mass and temperature are similar to the sources of 
\citet{elbadry2021b}. They are all in the transition from mass transfer to 
detached. Compared with the sources in \citet{elbadry2021b}, 
J0419 has the smallest surface gravity (see Figure \ref{fig:teff_logg}), 
which is due to its far longer orbital period. 
According to the evolution model of 
ELMs \citep{sun2018,Li2019}, pre-ELMs with longer initial periods will be 
more evolved before the mass transfer begins, resulting in smaller $\log g$ 
and longer periods. Long-period pre-ELMs like J0419 are rarely reported in 
previous works. 

Similar to \citet{elbadry2021b,Badry2021}, we show the position of J0419 on 
the HR diagram in Figure \ref{fig:HR}. For comparison, we plot the ELMs 
obtained from \citet{brown2020} and the pre-ELM objects obtained from 
 \citet{elbadry2021b}. We also show the CV sample using the data from 
 \citet{Ritter2003,Knigge2006}. The solid line in Figure \ref{fig:HR} is 
 the main sequence obtained from \texttt{isochrones} 
 \citep{isochrones2015} with an age of $\rm \log age = 8.3$.
Most CVs in Figure \ref{fig:HR} are fall on the main sequence.
Both the ELMs and pre-ELMs are well below the main sequence, showing that 
they are evolved stripped stars. 

\begin{figure*}[htp]
    \centering
    \includegraphics[width=0.7\textwidth]{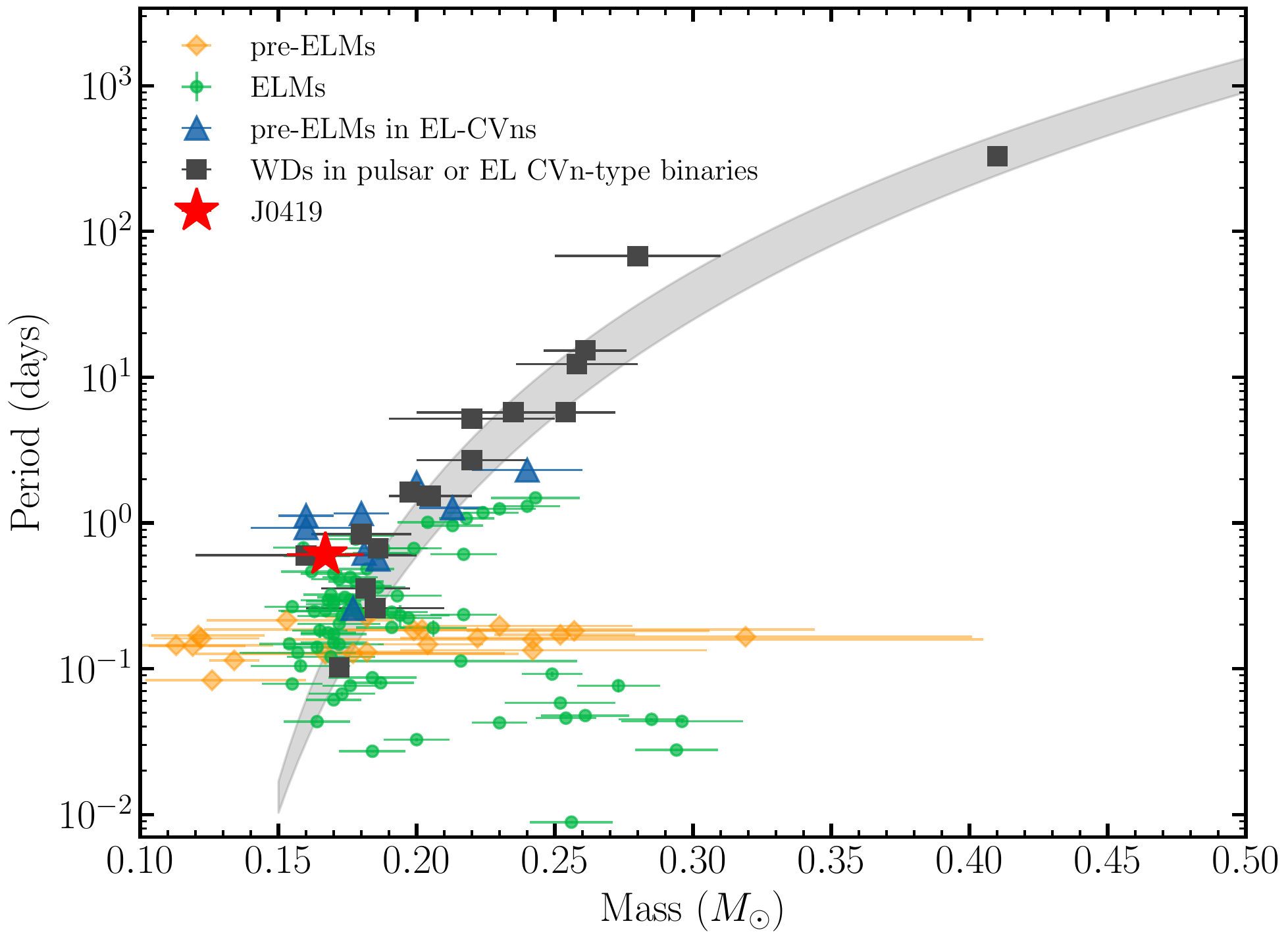}
    \caption{$M_\mathrm{WD}-P_\mathrm{orb}$ relation. 
    The black squares are helium WDs orbiting pulsars or pre-WDs orbiting an 
    A-type MS star \citep[see][]{Tauris2014}, and the data are obtained from 
    \citet{Antoniadis2012,Antoniadis2013,van2000,van2005b,van2010,Maxted2013,Corongiu2012,Jacoby2005,Ransom2014,Breton2012,Verbiest2008,Splaver2005,Pietrzy2012}. The blue triangles are pre-ELMs in EL CVns referred to 
    Section \ref{sec:cmp_pre_ELMs}. The green circle points are ELMs 
    reported in 
    \citet{brown2020}. The orange diamonds points are pre-ELMs from 
    \citet{elbadry2021b}. The red star is J0419. The shadow is the 
    $M_\mathrm{WD}-P_\mathrm{orb}$ relation calculated according to the 
    analytical formula in \citet{Tauris1999}. The upper and lower limits 
    of the shadow correspond to the metallicities of $\rm Z = 0.001-0.02$.
    }
    \label{fig:mass_period}
\end{figure*}

While the 
pre-ELMs of \citet{elbadry2021b} deviate significantly from the main 
sequence, J0419 almost falls on the main sequence, which makes the method 
of selecting J0419 analogs via HR diagram ineffective. For the J0419 analogs, 
we can only search for these objects by combining the 
variation features along with the SED fitting and spectroscopic information, 
which significantly limits their sample size. The multiple exposure strategy 
of LAMOST is beneficial to search for such long-period pre-ELMs and binary 
systems consisting of a visible star and a compact star \citep{yi2019}.

\subsection{The evolutionary properties}
\label{subsec:evo_prop}

According to \citet{sun2018,Li2019}, the orbit of J0419 will continue to 
shrink with angular momentum loss due to 
the magnetic braking until the convective envelope becomes too thin.
After the orbital contraction ends, the accretion process will 
stop, and the radius of the visible star begins to shrink with increasing 
temperature and $\log g$.  The visible star will gradually evolve into an 
ELM WD.

The typical temperature of the transition from mass-transferring CVs to 
detached ELMs is about 6500\,K \citep{sun2018}, which is related to the 
\textit{Kraft} break \citep{Kraft1967}. When the temperature is higher than 
this value, stars will lack the convective envelopes to generate the 
magnetic field, so that the magnetic braking is no longer effective. The 
sample of \citet{elbadry2021b} well verifies this statement.
In their sample, the emission lines only occur in the sources with 
$T_\mathrm{eff} < 6600$\,K, and the sources with higher temperatures have no 
emission lines found. J0419 with the donor temperature of 
$T_\mathrm{eff} = 5793$\,K seems also obey the law. 

The evolution of ELMs mainly depends on the initial mass of the visible star 
and the initial orbital period. The stars in WD + MS binaries with 
sufficiently long initial periods will ascend to the red giant branch when 
the mass transfer begins. The orbits of such systems will expand with the 
mass transfer (above the bifurcation period, see Figure 6 of 
\citealt{Li2019}). For the donors just leaving the main sequence 
when mass transfer begins, their orbits will shrink due to the magnetic 
braking. Both the mass and period of J0419 appear to be in between these 
two cases.

For WDs in binaries beyond the bifurcation period, there is a tight 
relationship between the core mass of a low mass giant and the radius of 
its envelope, resulting in a good correlation between the period and the 
core mass at the termination of mass transfer \citep{Rappaport1995}.
For systems below the bifurcation period 
($P_\mathrm{orb} \lesssim 16-22$\,h, 
$M_\mathrm{core} \lesssim 0.18 M_\odot$), the correlation between the radius 
and mass of the donors is unclear.

Figure \ref{fig:mass_period} shows the mass---period distribution 
of helium stars. The long-period systems are radio pulsar binaries 
or EL CVn-type systems, and most of the short-period objects are 
ELM/pre-ELMs reported in recent years. J0419 is just at the junction of 
the upper and the lower systems. With a mass of 
$M_1 = 0.176\pm0.014 M_\odot$ and a period of 
$P_\mathrm{orb} = 0.607189$~days, J0419 appears to follow the 
$M_\mathrm{WD} - P_\mathrm{orb}$ relation. 
The proprieties of closing to the bifurcation period and ongoing 
mass transfer make J0419 a unique source to connect ELM/pre-ELM 
systems, wide binaries, and CVs.
The systems with periods longer than 14 hours well follow the 
$M_\mathrm{WD} - P_\mathrm{orb}$ relation. But for short period targets, 
the correlation becomes diffuse. The sources with short periods and 
relatively large mass at the bottom of Figure \ref{fig:mass_period} are 
thought to be generated through CE channel \citep{Li2019}. 

\section{Summary}
\label{sec:summary}

We report a pre-ELM, J0419, consisting of a visible 
star and a compact star selected from the LAMOST medium-resolution survey with a period of 
$P_\mathrm{orb} = 0.607189$~days. The follow-up spectroscopic observations 
provide a RV semi-amplitude of the visible star, 
$K_1 = 216\pm 3$\,\kmps, 
yielding a mass function of 
$f(M_2) = 0.63 \pm 0.03 M_\odot$.
Both the large-amplitude ellipsoidal variability and the emission lines in 
the spectra indicate that the visible star has filled the Roche lobe. 
We use the mean density of the Roche lobe (only depending on the orbital 
period) and the radius of the visible star obtained from the SED fitting to 
calculate the mass of the visible star, which yields a mass of 
$M_1 = 0.176\pm0.014\,M_\odot$. The visible star mass is significantly
lower than the expected mass of a star with a G-type spectrum, indicating 
that the donor of J0419 is an evolved star, i.e., a pre-ELM. 
By fitting both the light and RV curves using 
\texttt{Phoebe}, we obtain an inclination angle of 
$i = 66.5_{-1.7}^{+1.4}$ degrees, corresponding to the compact 
object mass of $M_2 = 1.09\pm 0.05\,M_\odot$.
We find that 
J0419 has many features that are similar to the pre-ELM sample in 
\citet{elbadry2021b}, such as the visible star mass, temperature, the 
low mass transfer rate. However, the orbital 
period of J0419 is about three times the mean period of the sample in \citet{elbadry2021b}. 
We list the main 
properties of J0419 below:

\begin{itemize}
    
    \item
    J0419 
    has a more evolved donor than the objects of \citet{elbadry2021b}.
    The surface gravity of the visible star, $\log g = 3.9$, is smaller 
    than known pre-ELM systems, showing that the visible star of J0419 is 
    very bloated (see Section 
    \ref{sec:cmp_pre_ELMs} and Figure \ref{fig:teff_logg}).

    \item J0419 shows clear signatures of mass transfer. With a temperature 
    of $T_\mathrm{eff} = 5793_{-133}^{+124}$K, J0419 seems to obey the 
    empirical relation in \citet{elbadry2021b} that the low-temperature 
    pre-ELMs ($T_\mathrm{eff} < 6500 \sim 7000$\,K) have mass 
    transfer, but the pre-ELMs with higher temperature have no mass transfer. 
    The phenomenon
    is generally considered to 
    be caused by the magnetic braking becoming inefficient to take away the 
    angular momentum, and the donors shrink inside
    their Roche lobe (see Section \ref{subsec:evo_prop}). 

    \item According to the 
    evolutionary model of \citet{Li2019}, we suspect that J0419 may be a 
    rare source close to the bifurcation period of orbit evolution and 
    therefore did not shrink or expand its orbit significantly (see 
    Section \ref{subsec:evo_prop} and Figure \ref{fig:mass_period}). 

    \item J0419 is close to the main sequence, which makes the selection of the long-period pre-ELMs like J0419
    based on the HR diagram inefficient. Our work demonstrates a unique way to select such pre-ELMs by 
    combining time-domain photometric and spectroscopic observations 
    (see Figure \ref{fig:HR}).

\end{itemize}

\section{Acknowledgements}

We thank Fan Yang, Honggang Yang, and Weikai Zong for 
beneficial discussions, and thank the anonymous referee for constructive 
suggestions that improved the paper.
This work was supported by
the National Key R\&D Program of China under grant 2021YFA1600401,
and the National Natural Science Foundation of China under grants 11925301, 
12103041, 12033006, 11973002, 11988101, 11933004, 12090044, 11833006, 
U1831205, and U1938105.
This paper uses the LAMOST spectra. We also acknowledge the support of 
the staff of the Xinglong 2.16-meter telescope and Lijiang 2.4-meter 
telescope.

\software{IRAF \citep{Tody1986,Tody1993}, PyAstronomy \citep{Czesla2019}, 
lightkurve \citep{Lightkurve2018}, PyHammer \citep{Kesseli2017,Roulston2020}, The\_Payne \citep{Ting2019}, astroARIADNE \citep{Vines2022}, isochrones \citep{isochrones2015}, Phoebe \citep{prsa2005,prsa2016,Conroy2020}}

\bibliography{sample631}{}
\bibliographystyle{aasjournal}

\end{document}